\documentclass{elsart}

\usepackage[square,comma]{natbib}
\usepackage{graphicx}
\usepackage{pxfonts}
\usepackage{lineno}

\usepackage{amssymb}

\journal{}

\begin{document}

\thispagestyle{empty}
\begin{Large}
\textbf{DEUTSCHES ELEKTRONEN-SYNCHROTRON}

\textbf{\large{Ein Forschungszentrum der
Helmholtz-Gemeinschaft}\\}
\end{Large}

DESY 11-049

March 2011

\begin{eqnarray}
\nonumber &&\cr \nonumber && \cr \nonumber &&\cr
\end{eqnarray}
\begin{eqnarray}
\nonumber
\end{eqnarray}
\begin{center}
\begin{Large}
\textbf{Self-seeding scheme with gas monochromator for narrow-bandwidth soft
X-ray FELs}
\end{Large}
\begin{eqnarray}
\nonumber &&\cr \nonumber && \cr
\end{eqnarray}

\begin{large}
Gianluca Geloni,
\end{large}
\textsl{\\European XFEL GmbH, Hamburg}
\begin{large}

Vitali Kocharyan and Evgeni Saldin
\end{large}
\textsl{\\Deutsches Elektronen-Synchrotron DESY, Hamburg}
\begin{eqnarray}
\nonumber
\end{eqnarray}
\begin{eqnarray}
\nonumber
\end{eqnarray}
ISSN 0418-9833
\begin{eqnarray}
\nonumber
\end{eqnarray}
\begin{large}
\textbf{NOTKESTRASSE 85 - 22607 HAMBURG}
\end{large}
\end{center}
\clearpage
\newpage

\begin{frontmatter}



\title{Self-seeding scheme with gas monochromator for narrow-bandwidth soft
X-ray FELs}


\author[XFEL]{Gianluca Geloni\thanksref{corr},}
\thanks[corr]{Corresponding Author. E-mail address: gianluca.geloni@xfel.eu}
\author[DESY]{Vitali Kocharyan}
\author[DESY]{and Evgeni Saldin}

\address[XFEL]{European XFEL GmbH, Hamburg, Germany}
\address[DESY]{Deutsches Elektronen-Synchrotron (DESY), Hamburg,
Germany}

\begin{abstract}
Self-seeding schemes, consisting of two undulators with a monochromator in between, aim at reducing the bandwidth of SASE X-ray FELs. We recently proposed to use a new method of monochromatization exploiting a single crystal in Bragg-transmission geometry for self-seeding in the hard X-ray range. Here we consider a possible extension of this method to the soft X-ray range using a cell filled with resonantly absorbing gas as monochromator. The transmittance spectrum in the gas exhibits an absorbing resonance with narrow bandwidth. Then, similarly to the hard X-ray case, the temporal waveform of the transmitted radiation pulse is characterized by a long monochromatic wake. In fact, the FEL pulse forces the gas atoms to oscillate in a way consistent with a forward-propagating, monochromatic radiation beam. The radiation power within this wake is much larger than the equivalent shot noise power in the electron bunch. Further on, the monochromatic wake of the radiation pulse is combined with the delayed electron bunch and amplified in the second undulator. The proposed setup is extremely simple, and composed of as few as two simple elements. These are the gas cell, to be filled with noble gas, and a short magnetic chicane.  The installation of the magnetic chicane does not perturb the undulator focusing system and does not interfere with the baseline mode of operation.  In this paper we assess the features of gas monochromator based on the use of He and Ne. We analyze
the processes in the monochromator gas cell and outside it, touching upon the performance of the differential pumping system as well. We study the feasibility of using the proposed self-seeding technique to generate narrow bandwidth soft X-ray radiation in the LCLS-II soft X-ray beam line. We present conceptual design, technical implementation and expected performances of the gas monochromator self-seeding scheme.
\end{abstract}

%
%
%
\end{frontmatter}



\section{\label{sec:intro} Introduction}

Self-seeding schemes have been studied to reduce the bandwidth of SASE X-ray FEL \cite{SELF} - \cite{HUAN}. In general, a self-seeded FEL consists of two undulators with an monochromator located between them. The first undulator operates in the high gain linear regime starting from the shot noise in the electron beam. After the first undulator, the output radiation passes through the monochromator, which reduces the bandwidth to the desired value, smaller than the FEL bandwidth. While the radiation is sent through the monochromator, the electron beam passes through a bypass, which removes the electron microbunching introduced in the first undulator and compensates for the path delay created during the passage in the monochromator. At the entrance of the second undulator, the monochromatic radiation pulse is recombined with the electron beam, and amplified up to saturation. The radiation power at the entrance of the second undulator is dominant over the equivalent shot noise power, so that the bandwidth of the input signal is smaller than the bandwidth of FEL amplifier.

The self-seeding scheme proposed for the VUV and soft X-ray region makes use of a grating monochromator \cite{SELF} . The presence of such monochromator introduces a path delay with respect to the straight path, which has to be compensated with the introduction of a bypass of the length of about $30$ m for the case of the LCLS-II \cite{CDR2}. This requires relevant modifications of the baseline. We recently proposed to use a new method of monochromatization exploiting a single crystal in Bragg-transmission geometry for self-seeding in the hard X-ray range  \cite{OURX}-\cite{OURY6}. A great advantage of such transmission geometry method is that it introduces no path-delay of X-rays in the monochromator,  thus avoiding the need for a long electron bypass.

\begin{figure}
\includegraphics[width=1.0\textwidth]{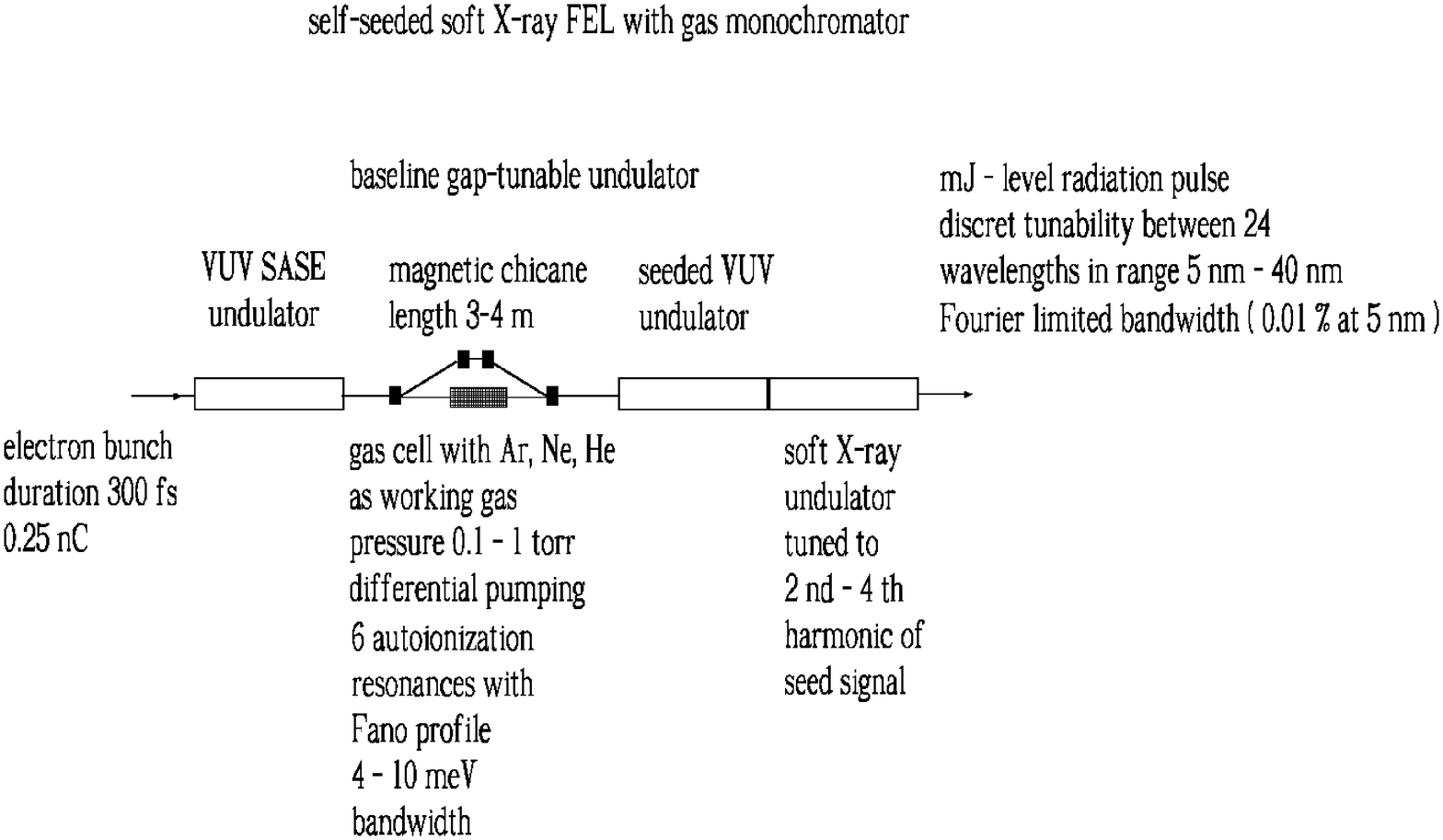}
\caption{Design of the VUV - soft X-ray FEL undulator system for the narrow
bandwidth mode of operation. The scheme is based on the use of the self-seeding scheme with a gas monochromator. The magnetic chicane absolves three tasks. It creates an offset for installing the gas monochromator, it removes the electron micro-bunching produced in the first undulator, and it shifts the electron bunch onto the monochromatic wake. After the gas cell, the monochromatic wake of the radiation pulse is combined with the delayed electron bunch, and amplified in the downstream undulator. The downstream undulator itself consists of two parts: a first part, in which the monochromatic seed signal is amplified in the linear regime and a second part, in which the harmonic of the seed signal produced in the first part is amplified up to saturation.} \label{SX5}
\end{figure}
All VUV and soft X-ray monochromators operate in the frequency domain as bandpass filters. In this paper we propose, instead, a new method of monochromatization based on the use a cell containing resonantly absorbing gas in the transmission direction. Such cell works, in analogy with the single-crystal monochromator method for the hard X-ray, as a bandstop filter. Then, such problems as extra path-delay of radiation pulse, heat loading and monochromator alignment are solved. The proposed setup is extremely simple, and composed of two simple elements: a gas cell and a short magnetic chicane, Fig. \ref{SX5}. As for the hard X-ray case, the magnetic chicane accomplishes three tasks. It creates an offset for gas cell installation, it removes the electron micro bunching produced in the first undulator, and it acts as a delay line for the electron bunch. Thus, using a cell with rare gas installed within a short magnetic chicane in the baseline soft X-ray undulator as a bandstop filter, it is possible to decrease the  bandwidth of the VUV radiation down to the Fourier limit.

We will show that the proposed technique can be implemented in the soft X-ray region, exploiting the tunable-gap soft X-ray baseline undulator. We propose to first perform monochromatization at $20-40$ nm with the help of the gas-cell monochromator, and subsequently amplify the radiation in the first part the of output undulator. The amplification process can be stopped at some position well before the FEL reaches saturation, where the electron beam gets considerable bunching at the 2nd, 3rd and 4th harmonic of the coherent radiation. The undulator downstream that position can then be tuned to the harmonic frequency to amplify
the radiation further to saturation, Fig. \ref{SX5}.

In this paper we present a detailed study for applying the self-seeding scheme with gas monochromator to the LCLS-II setup, in order to generate narrow bandwidth, soft X-ray radiation in the proposed soft X-ray beam line.  The tentative design for the proposed technique at LCLS-II allows to generate fully coherent radiation at a wavelength of about $5$ nm.  We focus on the conceptual design, on the technical implementation and on the expected performances of the self-seeding scheme using the gas monochromator.

\section{\label{prin} Principles of the self-seeding technique based on the use of a gas
monochromator}

\begin{figure}
\includegraphics[width=1.0\textwidth]{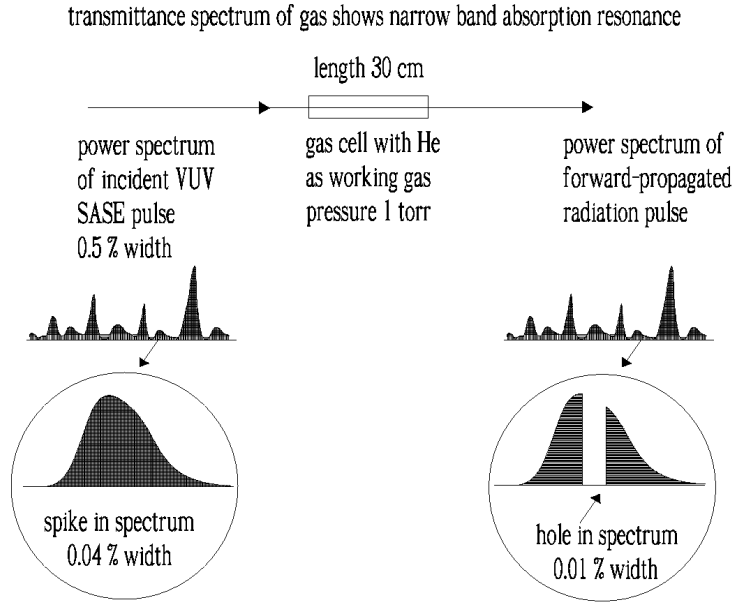}
\caption{Gas cell with He, Ne and Ar as working gas, working as a bandstop filter for the transmitted VUV SASE radiation pulse.} \label{SX3}
\end{figure}

\begin{figure}
\includegraphics[width=1.0\textwidth]{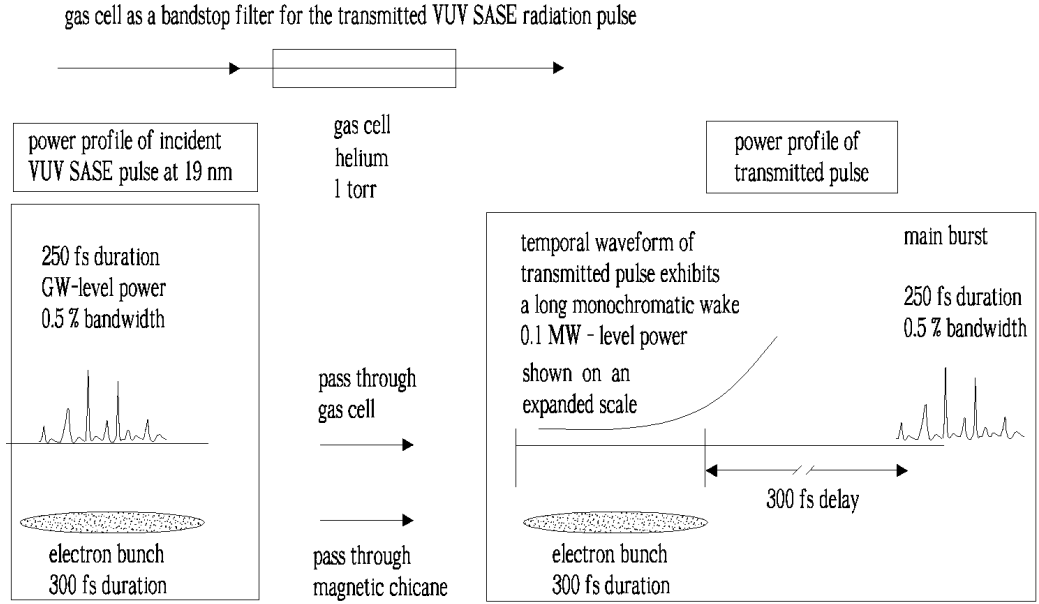}
\caption{Forward radiation in the gas cell. The transmittance spectrum in the gas cell shows an absorption resonance with narrow (4-10 meV) line width. When the spectral contents of the incoming radiation beam overlap the absorption resonance, the temporal waveform of the transmitted radiation pulse exhibits a long monochromatic wake. The duration of the wake is inversely proportional to the bandwidth of the absorption resonance.} \label{SX4}
\end{figure}

In this section we illustrate more in detail our novel method of monochromatization, based on the use of a cell containing resonantly absorbing gas. As already said, this technique takes advantage  of the transmission geometry, where no extra path-delay for the radiation pulse is present. The principle of the new method of monochromatization is very simple and is illustrated in Fig. \ref{SX3} and Fig. \ref{SX4}. An incident SASE pulse coming from the first undulator impinges on the gas
cell. When the spectral contents of the incoming SASE  beam satisfies the photo-absorption resonance condition of the gas, the gas cell actually operates as a bandstop filter for the transmitted VUV SASE radiation pulse, see Fig. \ref{SX4}.

Obviously, if we use a bandstop filter there is no monochromatization in the frequency domain. However, the temporal waveform of the transmitted radiation pulse shows a long monochromatic wake. The duration of this wake is inversely proportional to the bandwidth of the absorption line in the transmittance spectrum. It is then possible to reach a bandwidth limited seed signal by exploiting a temporal windowing technique, as indicated in Fig. \ref{SX4}. Since we deal with a high-gain parametric amplifier where the properties of the active medium, i.e. electron beam, depend on time, the temporal windowing concept can be practically
implemented in a simple way by delaying the electron bunch at the position where the frequency spectrum of the transmitted pulse experiences a strong temporal separation. In other words, the magnetic chicane in Fig. \ref{SX5} shifts the electron bunch on top of the monochromatic wake created by the bandstop filter. By this, it is possible to seed the electron bunch with radiation pulse characterized by bandwidth much narrower than the natural FEL bandwidth. The concept is similar to that developed in \cite{OURX}-\cite{OURY6} for the hard X-ray case. However, in the proposed scheme no  optical elements are used, and problems with alignment or heat loading do not exist at all. While the radiation is sent through the gas cell, the electron beam passes through the magnetic chicane, which accomplishes three tasks by itself: it creates an offset for the gas cell installation, it removes the electron micro bunching produced in the first (SASE) undulator and, as already said, it acts as a delay line for the implementation of the temporal windowing.

For the VUV wavelength range, a momentum compaction factor $R_{56}$ in the order of $100~\mu$m is sufficient to remove the microbunching in the electron beam. As a result, the choice of the strength of the magnetic chicane only depends on the delay that we need to introduce between electrons and radiation. In our case, this amounts to $100~\mu$m, see Fig. \ref{SX5}. Such delay is small enough to be generated by a short $3-4$ m long chicane to be installed in place of a single undulator module at a facility like the LCLS-II. Such chicane is, however, strong enough to create a sufficiently large transverse offset of $1-2$ cm for installing the gas cell.

\section{\label{phab} VUV Photo-absorption spectra of the rare gases}

In this article we will take advantage of autoionizing resonances in rare gases. We consider He, Ne and Ar as possible choices,  presenting  cross sections and tables of line parameters. Additionally we will discuss about Doppler broadening as an effect, which can be accounted for in this most general level of description\footnote{The Doppler broadening depends on the gas temperature only (room temperature in our case) and not on density or pressure, which will be specified only later, when discussing the differential pumping system design.}.

The phenomenon of autoionization is well-known in literature (see for example textbooks like \cite{BOHM}, \cite{FANO} and references therein, just as examples of reviews). Let us consider the Helium atom as a case study. The interaction with a soft X-ray electromagnetic pulse has the effect of ionizing the atom.  A first ionization threshold, where the remaining electron in the ion is found in the state with principal quantum number $N=1$, is found at $24.6$ eV, while a second ionization threshold with $N=2$ is found at $65.4$ eV. Therefore, after interaction with a photon with energy between these two thresholds,  one finds the Helium ion in the ground state $1s$, and a free electron with kinetic energy equal to the photon energy diminished of  $24.6$ eV. This single open channel can be reached either directly or via decay of a doubly excited state, called autoionizing state. A set of autoionizing states can be excited at different energies. If the photon energy increase between $65.4$ eV to $72.9$ eV, not only the previously discussed open channel is possible. In fact, above the second ionization threshold a second open channel is present, characterized by  the Helium ion with the bound electron with $N=2$ ($2s$ state), and free electron energy equal to the photon energy diminished by $65.4$ eV. In this photon energy range we have a set of autoionizing resonances,which interfere with two open channels ($1s$ or $2s$). More in general, as the photon energy increases below the double ionization threshold at $79$ eV, other channels corresponding to the Helium ion with higher numbers of $N$ ($N$th ionization threshold) can be reached.

The autoionizing $He^*$ state corresponds to a double electronic excitation where the two electron principal numbers become $n_1 = N$ and $n_2 =N, N+1, ...$, with $N>1$. Each choice of $N=2,3...$ corresponds to a set of Rydberg series. Dipole transitions allow $2N+1$ series of  $~^1 P_0$ autoionizing $He^*$ states. These series of resonances admit as a limiting energy the $N$th ionization threshold corresponding to the an excited state with quantum number $N$ for the $He^+$ ion. Since for a given choice $N$ series admits the $N$th ionization threshold as a limit, this means that any autoionizing $He^*$ state decays to the $He^+$ state with quantum number $N-1$.

In this paper we will focus on the series of autoionizing resonances converging to the $N = 2$ levels of Helium ion. This means that we have $2N+1 = 3$ Rydberg series, all converging to the second ionization threshold, corresponding to $65.4$ eV. These three series all begin with the two Helium electrons characterized by $n_1 = 2$ and $n_2 = 2$, which corresponds to an energy of $60$ eV. There are three possible choices for the $(l_1,l_2)$ quantum numbers: $(2s,np)~^1P_0$, $(2p,ns)~^1P_0$, and $(2p,nd)~^1P_0$. However, the levels of the $(2s,np)~^1P_0$ and $(2p,ns)~^1P_0$ wave functions are nearly degenerate, and the doubly-excited states are better described in terms of constructive and destructive superpositions of these wave functions, finally yielding the three Rydberg series: $(sp,2n+)~^1P_0$, $(sp,2n-)~^1P_0$, and $(2p,nd)~^1P_0$. Since these three series converge (but no reach) the second ionization threshold, all the correspondent doubly excited states $He^*$ decay to an Helium ion in the ground state. Out of these three series, only the first $(sp,2n+)$ will be of interest to us, as the others are too weak.

\subsection{Photo-absorption cross-section of He for photon energies 60 eV to 65 eV}

As we just said, autoionizing resonances results from the decay of doubly excited Rydberg states of $He^*$ into the continuum, i.e. $He^* \longrightarrow He^+  e^{-}$. Since the continuum can also be reached by direct photoionization, both paths add coherently, giving rise to an interference. This interference is related with the typical Fano line shape for the cross-section as a function of energy \cite{FANO2}. A more detailed physical explanation of the entire process of interaction of the FEL beam with the gas is given in Section \ref{six} and \ref{seven}. Here we limit to report the cross section for these series, which can be modeled by the expression \cite{MORG}

\begin{eqnarray}
\sigma(\lambda,q,\Gamma) = \sigma_b(\lambda)\left(\frac{\left(\sum_{n=2}^{\infty}(q_n/\mathcal{E}_n)+1\right)^2}{\sum_{n=2}^\infty(1/\mathcal{E}_n)^2 +1}\right)~,
\label{sigsig}
\end{eqnarray}
where the energy-dependent background cross-section expressed in Megabarn ($1Mb = 10^{-18} cm^{2}$) is given by

\begin{eqnarray}
\sigma_b(\lambda) = -0.05504-1.3624\cdot 10^{-4} \lambda + 3.3822\cdot 10^{-5} \lambda^2~,
\label{sigb}
\end{eqnarray}
with $\lambda$ the radiation wavelength in Angstrom units, while the reduced energy $\mathcal{E}_n$ is defined as

\begin{eqnarray}
\mathcal{E}_n = \frac{2(E_{R_n}-hc/\lambda)}{\Gamma_n}~.
\label{En}
\end{eqnarray}
The asymmetry index $q_n$, the energy of the $n$th resonance $E_{R_n}$, the resonance width $\Gamma_n$, have been the subject of several calculations. \begin{table}
  \caption{Parameters determining the Fano autoionization profiles for the  $(sp,2n+)$ autoionizing series of Helium, following \cite{MORG}.}
{\begin{tabular}[l]{@{}|c||c|c|c|c|c|c|c|c|c|}

\hline
    n              & 2      &3      &4      &5      &6      &7      &8      &9      &10          \\
    $\Gamma$  (eV) & 0.0378 & 0.0083& 0.0038& 0.0014& 0.0008& 0.0005& 0.0003& 0.0002& 0.0001     \\
    $E_{res}$ (eV) &60.14   & 63.655& 64.466& 64.816& 64.999& 65.108& 65.181& 65.229& 65.263     \\
    q              & -2.57  & -2.5  & -2.5  & -2.5  & -2.5  & -2.5  & -2.5  & - 2.5 & -2.5       \\
\hline
  \end{tabular}}
  \label{tableHe}
\end{table}

\begin{figure}
\includegraphics[width=1.0\textwidth]{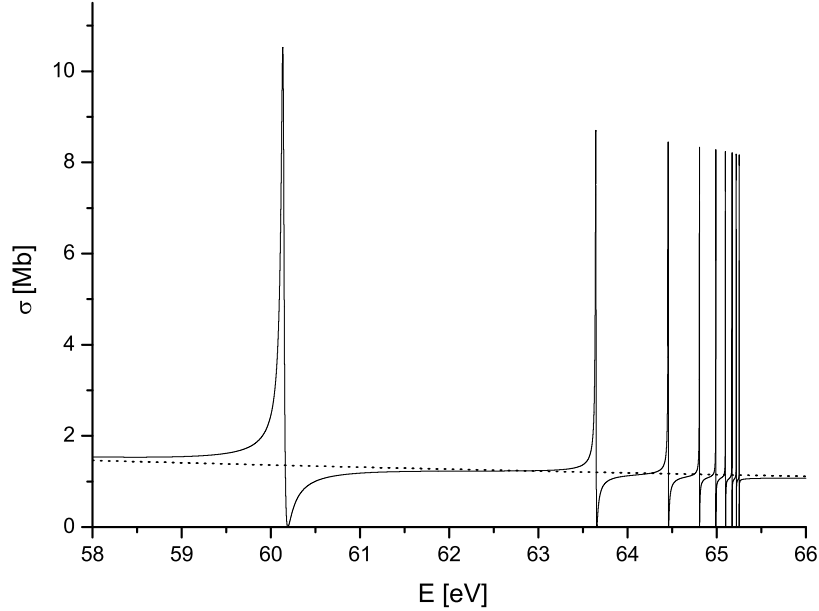}
\caption{Fano profiles for the $(sp,2n+)$ autoionizing series of Helium. The cross-sections are calculated following \cite{MORG}.
} \label{crossec}
\end{figure}
The cross-section for the series $(sp,2n+)$ can be calculated using Eq. (\ref{sigsig}), with the parameters given in Table \ref{tableHe}, reproduced from \cite{MORG}. Also, the photoabsorption cross section $\sigma$ is linked to the light attenuation through a gas medium of column density $n_0 l$, $l$ being the length of the cell and $n_0$ the gas density. This allows cross-section measurements via the Beer-Lambert law,

\begin{eqnarray}
I(\omega) = I_0 exp[-n_0 l \sigma(\omega)],
\label{beerl}
\end{eqnarray}
where $I_0$ is the incident intensity, and $I(\omega)$ is the attenuated intensity of the transmitted light at frequency $\omega$. The values of parameters obtained from experiments are in general consistent with calculations these \cite{MORG}.

The resulting cross-section for the  $(sp,2n+)$ autoionizing series of Helium is shown in Fig. \ref{crossec} as a function of the energy. The availability of a series resonances allow for tunability of the scheme. However, not all resonances can be used. In fact, they should be located further away from each other than the FEL bandwidth, and be narrow enough to guarantee the application of the temporal windowing process. In our case, a rule of thumb requires a width narrower than $10$ meV. In this case, only the second and the third Helium lines are suitable, as the first one is too wide, and starting from the fourth there is not enough separation\footnote{Such smaller separation maybe useful at VUV facilities, in order to seed at multiple photon wavelengths with separation of a fraction of eV, similarly to what has been proposed in \cite{OURY6}.}.

\subsection{Photo-absorption cross-section of Ne for photon energies 45-49 eV}

In this paper we also study of possibility to use the autoionizing resonances of Neon from $2s^2 2p^6$ to $2s 2p^6 np$ , with $n = 3-5$ \cite{NEON}. This allows our setup for further tunability. It should be noted that there is a difference in the calculation of the cross-section, when compared with the Helium case for energies below the second ionization threshold. In fact, in that case only one channel is open, meaning that the final state is the Helium ion in the ground state. The two paths towards this channels interfere and one is left with the Fano line considered before. However, when the energy increases, or one considers another gas like Neon, other channels open up. Then, as explained for example in \cite{ARGO}, transitions to the continuum may or may not interact with the discrete autoionization state, and the expression for the cross-section has to be modified as:

\begin{eqnarray}
\sigma = \sigma_b \frac{(q+\mathcal{E})^2}{1+\mathcal{E}^2} + \sigma_a~,
\label{sigsig2}
\end{eqnarray}
where $\sigma_b$ and $\sigma_a$ are the cross-sections of the interacting and non-interacting transitions to the continuum. Defining the total cross-section $\sigma_0= \sigma_a+\sigma_b$, which one would measure without autoionization, Eq. (\ref{sigsig2}) can be cast in the form

\begin{eqnarray}
\sigma = \sigma_0 \left(\rho^2 \frac{(q+\mathcal{E})^2}{1+\mathcal{E}^2} -\rho^2+1 \right)~,
\label{sigsig3}
\end{eqnarray}
where we also defined the correlation coefficient (equal to unity for the Helium resonances under study):

\begin{eqnarray}
\rho^2 = \frac{\sigma_b}{\sigma_b+\sigma_a}~.
\label{rho2}
\end{eqnarray}
The relevant resonances can be calculated using Eq. (\ref{sigsig3}), with the parameters given in Table \ref{tableNe}, reproduced from \cite{NEON}. 
All three resonances look interesting for the implementation of our technique, although the one corresponding to $n=3$ is somewhat larger than $10$ meV. It should also be remarked that the presence of a correlation parameter different from unity leads to a fundamental difference in the Fano profiles, that is, the cross-section never goes to zero, at variance with the Helium case.

\begin{table}
  \caption{Parameters determining the Fano autoionization profiles for the  $(sp,2n+)$ autoionizing series of Helium, following \cite{MORG}.}
{\begin{tabular}[l]{@{}|c||c|c|c|}

\hline
    n              & 3      &4      &5          \\
    $\sigma_0$ (Mb)& 8.6    & 8.0   & 8.2       \\
    $\Gamma$  (eV) & 0.013  & 0.0045& 0.0002    \\
    $E_{res}$ (eV) &45.546  & 47.121& 47.692    \\
    q              & -1.6   & -1.6  & -1.6      \\
    $\rho^2$       & 0.7    & 0.7   & 0.7       \\
\hline
  \end{tabular}}
  \label{tableNe}
\end{table}

\subsection{Photo-absorption cross-section of Ar for photon energy equal to 28.5 eV}

Another possible choice \cite{ARGO}, suitable for our purposes would be the resonance of the Argon gas at $28.5$ eV. Such resonance can also be calculated with the help of Eq. (\ref{sigsig3}), and is characterized by $\rho^2 = 0.85$, $q=0.17$ and has a width $\Gamma = 12.6$ meV.

\subsection{Doppler broadening}

To conclude the present Section we discuss Doppler broadening of the Fano lines. Due to the finite temperature $T$ of the gas, the frequency distribution of an ensemble of radiating atoms will experience a spread \cite{LODO}. Assuming the gas ensemble in thermal equilibrium, the Doppler shift is linked to the Maxwell-Boltzman probability distribution of velocities of the atoms. The distribution of frequencies turns out to be a Gaussian, with a peak at the resonance frequency and a FWHM given by

\begin{eqnarray}
\Gamma_D = 7 \cdot 10^{-7} E_0 \sqrt{T/M}~,
\label{doppler}
\end{eqnarray}
where $E_0$ is photon energy at resonance, $T$ is the temperature measured in Kelvin degrees, and $M$ is the molecular weight. Assuming a temperature of $300$ K, the Doppler broadening is pressure independent and turns out to amount to about $0.4$ meV for He ( molecular weight $M = 4$ ). For resonances under study, $\Gamma_D $is substantially smaller than the natural width $\Gamma \sim 4-10$ meV. It follows that simulations directly performed
from the calculated Fano profiles constitutes a good approximation of the exact profile, which should be obtained by convolving the Fano natural line with the Gaussian profile due to Doppler shift.

\section{\label{TTT} Transmittance of the gas cell used in monochromatization}

\begin{figure}
\includegraphics[width=1.0\textwidth]{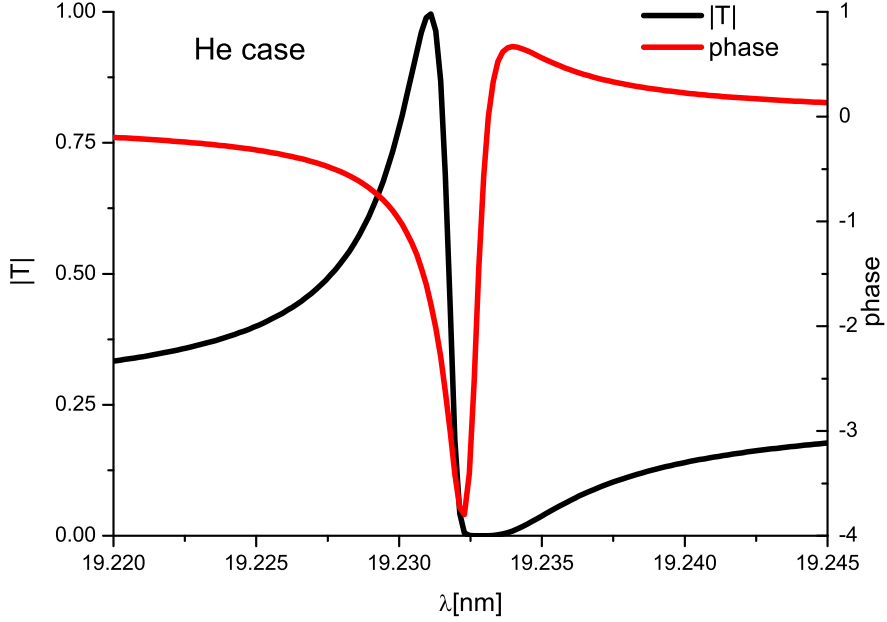}
\caption{Modulus and phase of the transmissivity of Helium around the $n=4$ line of the $(sp,2n+)~^1P_0$ Rydberg series. The modulus has been calculated according to Eq. (\ref{ModT}), while the phase is recovered with the help of the Kramers-Kroning relation according to Eq. (\ref{KKrel2}). } \label{transm}
\end{figure}

As discussed in Section \ref{phab}, if a monochromatic electromagnetic pulse of intensity $I_0$ and frequency $\omega$ impinges on a cell of length $l$, filled with a gas with density $n_0$, the transmitted intensity obeys the Beer-Lambert law, Eq. \ref{beerl}. As a result, by comparison with Eq. \ref{beerl}, the modulus of the transmissivity can be defined as

\begin{eqnarray}
|T| = \exp[-n_0 l \sigma/2]~,
\label{ModT}
\end{eqnarray}
where the shape of $\sigma(\omega)$, which follows a Fano profile, has been discussed in the previous Section, and the product $n_0 l$ is known as the column-density of the gas cell. In the following we will choose a column-density equal to  $n_0 l =  10^{18} cm^{-2}$, and we will restrict ourselves to the third ($n=4$) resonant line of the $(sp,2n+)~^1P_0$ Rydberg series for Helium, whose cross-sections can be calculated with the help of Eq. (\ref{sigsig}) and Eq. (\ref{sigb}).

Once $|T|$ is known, it is possible to use Kramers-Kroning relations to recover the phase.  In fact one can write

\begin{eqnarray}
\mathrm{ln}[T(\omega)] = \mathrm{ln}[|T(\omega)|] + i
\Phi(\omega) = - nl \sigma/2 + i
\Phi(\omega)~.\label{ln}
\end{eqnarray}
Note that $T^*(\omega)=T(-\omega)$ implies that $|T(\omega)|=|T(-\omega)|$ and that $\Phi(\omega) = - \Phi(-\omega)$. Therefore, using Eq. (\ref{ln}) one also has that $\mathrm{ln}[T(\omega)]^*=\mathrm{ln}[T(-\omega)]$. Then, application of Titchmarsh theorem shows that the analyticity of $\mathrm{ln}[|T(\Omega)|]$ on the upper complex $\Omega$-plane implies that

\begin{eqnarray}
\Phi(\omega)=-\frac{2\omega}{\pi}\mathcal{P} \int_0^{\infty}
\frac{\mathrm{ln}[T(\omega')] }{\omega'^2-\omega^2} d\omega'~,
\label{KKrel2}
\end{eqnarray}
A direct use of Eq. (\ref{KKrel2}), with $|T|$ given as in Eq. (\ref{ModT}), should yield back the phase $\Phi(\omega)$. As is well known one tacitly assumes that $\mathrm{ln}[T(\Omega)]$ is analytical on the upper complex $\Omega$-plane. This fact however is immediately granted by the fact that $\sigma$ is proportional to the imaginary part of the refractive index of the medium, and it is well known that the refractive index must obey Kramers-Kroning relation Eq. (\ref{KKrel2}). We therefore used Eq. (\ref{KKrel2}) in order to recover the phase of the transmittance. More specifically, we took advantage of a publicly available Matlab script \cite{LUC2} which serves exactly to that end. The final result in terms of modulus and phase of the transmissivity $T$ is shown in Fig. \ref{transm}.

\section{Physical analysis of the gas cell used in the monochromatization process}

\begin{figure}
\includegraphics[width=1.0\textwidth]{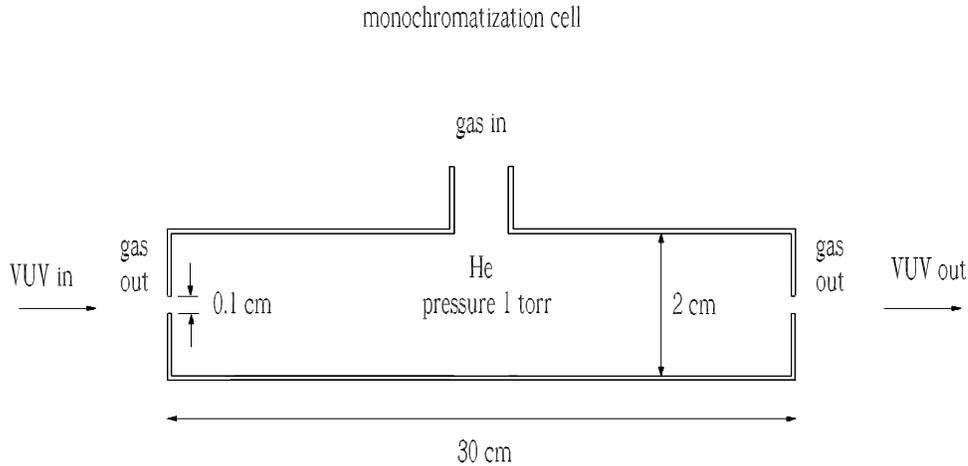}
\caption{Gas cell dimensions.
} \label{SX6}
\end{figure}
An analysis of the gas cell used in the monochromatization process is particularly important, since in the VUV range one cannot install a window separating the undulator pipe from the gas cell. The solution is based on the exploitation of a differential pumping scheme.  The analysis presented here is based on \cite{RYU1}-\cite{RYU5}, which where used in the design of the LCLS gas attenuator. Fig. \ref{SX6} shows the schematic of the gas cell that will be actually used in the monochromatization process. We will denote this cell as volume "0".

\begin{figure}
\includegraphics[width=1.0\textwidth]{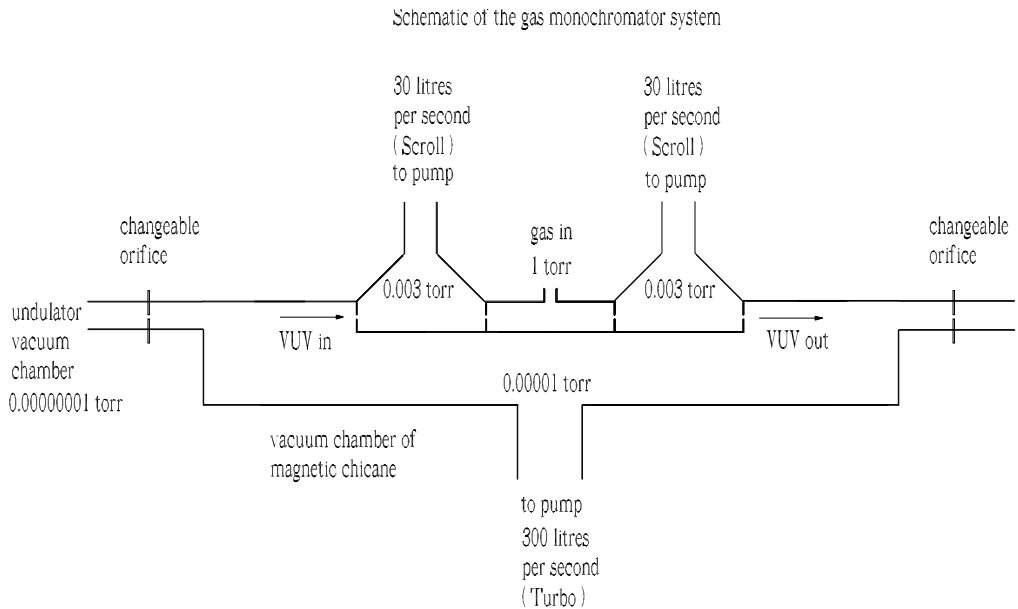}
\caption{Sketch of differential pumping system
} \label{SX7}
\end{figure}

Fig. \ref{SX7} shows how the gas cell is integrated in the differential pumping system. Two differential pumping cells are locates symmetrically around voulme "0",  after main cell (system symmetric in both sides). These cells will be denoted as volumes "1" and, similarly to volume "0", will have a length of $30$ cm and a radius of $1$ cm. They will be pumped by scroll vacuum pumps with a pumping speed of $30$ l/s. These three cells (volumes "1" and "0") are installed by exploiting the transverse offset given by the magnetic chicane, which is limited to a maximum of $1$ cm due to the requirement on the delay between electrons and X-rays. The system connected with vacuum chamber of magnetic chicane, which will be denoted as volume "2", through apertures of radius $a=0.5$ mm. At both ends of the vacuum chamber, the VUV are allowed to enter and exit through orifices with the same diameter. The dimensions of these apertures can obviously be changed. This means that if gas cell is not filled the orifices can be eliminated by fully opening them. When the gas cell operates, the orifices are in instead.   The vacuum chamber volume "2"  will be pumped by a turbo pump with a pumping speed of $300$ l/s. As a result we have differential pumping system which includes two stages per each side: one of the two volume "1" cells as first stage, and the volume "2" vacuum chamber as second stage. A gas outflow of $10^{14}$ atoms/s to the undulator vacuum chamber is obtainable, and is the same which was calculated for the LCLS gas attenuator, upon which the present analysis is based.

A technical problem to be overcome pertains the diameter of the vacuum pump port. For different pressures different kind of pumps will be used. As said above, the two volume "1" cells will be at a pressure of $0.003$ Torr. In this case scroll pumps will be used, with a port radius of $2$ cm. Volume "2" will be at a pressure of $0.00001$ Torr, in which case a turbo-pump will be employed, with a port radius of $5$ cm. For the vacuum chamber in the magnetic chicane, volume "2", the turbo pump can be operated without issues, but for the scroll pumps in the two volume "1" cells, one has to deal with a small transverse offset in the centimeter size, while the port radius of the pump is about $2$ cm. However, such constraint is only on the side where the beam offset is present.  On the other side, Fig. \ref{SX7}, an adiabatic geometrical transformer can be used connecting the side of volume "1" ($1$ cm radius and $30$ cm long) to the $2$ cm radius tube needed for the pump connection.

Let us assume that the gas used in the monochromator is Helium. For Neon, all issues associated with the gas monochromator become easier. In fact, due to the larger photo-absorption cross-section of Neon, the pressure in the cell can be an order of magnitude lower. Assuming that the gas in all systems is at roughly room temperature, the pressure $P$ and the gas density $n_0$ are related by

\begin{eqnarray}
n_0 [cm^{-3}] = 3.5 \cdot 10^{16} P [Torr]~.
\label{nP}
\end{eqnarray}
Given the length $l= 30$ cm of volume "0", that is the absorption cell,  one sees that a pressure $P_0 \sim 1$ Torr is needed to reach the column density $n_0 l = 10^{18}$ cm$^{-2}$ assumed in the previus sections.

The collisional mean free path $l_{col}$ is related to the density and to the elastic cross section $\sigma_e$ by $l_{col} = 1/(n_0 \sigma_e)$. Given the diameter of He atom $d_a  = 2.4 {\AA}$, the He-He elastic collision cross-section can be estimated as $\sigma_e \sim \pi (2 d_a)^2 \sim 7 \cdot 10^{-16}$ cm$^2$. It follows that, at the pressure of $1$ Torr the mean free path is $l_{col} \sim 5 \cdot 10^{-3}$ cm. This means that the gas flow is collisional, in the sense that the mean free path $l_{col}$ is much shorter than the orifice radius $a$. As a result, the gas exhaust through the orifices will occur in hydrodynamical manner (and not in Knudsen manner), and the number of atoms leaving the cell through each of the orifices can be
evaluated as

\begin{eqnarray}
Q_1 \sim 0.5 \pi a^2 n_0 s_g~,
\label{Q1}
\end{eqnarray}
where $s_g$ is speed of sound, and the multiplicative factor $0.5$ accounts to the reduction of the local speed of sound  and density at the transmission point. Taking the density in volume "0" $n_0 \sim 3\cdot 10^{16} cm^{-3}$, $s_g \sim 3 \cdot 10^4$ cm/s, and $a = 0.5$ mm one
can calculate the throughput as $Q_1 = 3 \cdot 10^{18}$ atoms/s.

The gas density in volumes "1" depends on the pump speed of the corresponding pumps. The gas-load equation reads

\begin{eqnarray}
Q_1 = S_p n_1 ~,
\label{Q1b}
\end{eqnarray}
where $S_p$ is the pumping speed in $l/s$, and $n_1$ is the gas density in volumes "1". Taking, as said above, the pumping speed equal to $S_p = 30 l/s$ and the throughput $Q_1 = 3 \cdot 10^{18}$ atoms/s, one can calculate the density and the pressure in cell "1" as as $n_1 \sim 10^{14} cm^{-3}$
and $P_1 \sim 3 \cdot 10^{-3}$ Torr.

Consider now the transition between cell "1" and volume "2". Here the mean-free path in the incoming gas is much larger than the orifice size $a = 0.5$ mm. The flow through the orifice under such circumstances is Knudsen flow. The vast majority of the atoms entering the volume "2"  from the cell  "1"  hits the walls of volume "2" and get scattered. These atoms are responsible for the density value $n_2$. The gas outflow is

\begin{eqnarray}
Q_2 \sim a^2 n_1 \sqrt{2\pi k T / M_{He}}
\label{Q2}
\end{eqnarray}
Using parameters $n_1 \sim 10^{14} \mathrm{cm}^{-3}$, $a = 0.5$ mm, and $T \sim 300$ K, this yields $Q_2 \sim 5 \cdot 10^{16}$ atoms/s. We assume a pump speed for volume "2" of $S_p \sim 300 l/s$, which  can be easily reached with various types of pumps, if the surface area of the vacuum port in the side walls
of volume is greater than $100$ cm$^2$. The density and pressure in volume "2" turns out to be then $n_2 \sim 2 \cdot 10^{11}$ cm$^{-3}$, and $P_2 \sim 10^{-5}$ Torr respectively.

We can now estimate the gas flow from the gas monochromator to the undulator beam line. The gas outflow to the undulator can be calculated in the same manner as the gas flow from the cell "1" to volume "2" and does not exceed

\begin{eqnarray}
Q_{un} \sim a^2 n_2 \sqrt{2 \pi k T/M_{He}} \sim 10^{14} ~\mathrm{atoms/s} .
\label{Qun}
\end{eqnarray}
or equivalently $Q_{un} \sim 3 \cdot 10^{-6}$ Torr l/s.  For a pumping speed in the undulator beamline $S_p \sim 300 l/s$, the final pressure
will then be $P_{un} \sim 10^{-8}$ Torr.

\section{\label{six} Theoretical background of line profiles in the VUV absorption spectra
of the rare gases}

In Section \ref{phab} we used the expression in Eq. (\ref{sigsig2}) for our cross-section calculations.  In this Section we give a justification of that expression based on \cite{COHE}. Let us consider for simplicity the Helium case below the second ionization threshold ($\sigma_a = 0$). The FEL radiation can be treated as a time-dependent perturbation of a two-state system, the first state $|1>$ corresponding to the Helium ground state, and the second state $|2>$ corresponding to the final ionized state. In this case (see, any textbook of Quantum Mechanics, for example \cite{GRIF}), we can make use of the Fermi's Golden Rule to calculate the transition rate between these states. Let us indicate with $<2|p|1>$ the matrix element of the perturbing potential, where the dipole moment operator $p$ is given by

\begin{eqnarray}
p = - e \sum_{j=1}^{N_e} r_j~,
\label{P}
\end{eqnarray}
$r_j$ being the transverse displacement of the $j$th electron in the atom. The transition rate can be found with the help of the Fermi's Golden Rule as \cite{GRIF}:

\begin{eqnarray}
W_{12} = \frac{\pi}{\hbar^2} \sqrt{\frac{\mu_0}{\epsilon_0}} |<2|p|1>|^2 I(\omega_{12})
\label{gold}
\end{eqnarray}
where $I$ is the intensity, and $\omega_{12}$ is the transition frequency. A relation between absorbed power $P$ and incoming intensity $I$ can be found by multiplying both ends by the photon energy $\hbar \omega_{12}$, yielding

\begin{eqnarray}
P = \frac{\pi}{\hbar} \sqrt{\frac{\mu_0}{\epsilon_0}} |<2|p|1>|^2 \omega_{12} I(\omega_{12})
\label{gold2}
\end{eqnarray}
Since by definition the cross section is the ration between absorbed power $P$ and incoming intensity $I$, it follows that the cross section is proportional to the square of the transmission amplitude.  In order to calculate the cross-section in Eq. (\ref{sigsig2}) we therefore need to calculate the square of the transmission amplitude.

The interaction with the electromagnetic pulse has the effect of ionizing the atom. As discussed in Section \ref{phab}, at some particular energies, ionization can happen either directly: $He + \hbar \omega \longrightarrow He^+ + e^-$ or through a discrete autoionization state $He^*$ : $He + \hbar \omega \longrightarrow He^* \longrightarrow  He^+ + e^-$. In the example case of Helium below the second ionization threshold considered here, such particular state $He^*$ corresponds to a double electronic excitation where the two electron principal numbers become $n_1 = 2$ and $n_2 =2, 3...$.

In the quantum-mechanical language, what was just said means that the electromagnetic perturbation couples the ground state $|\psi_0>$ to the quasi-continuum states\footnote{Following \cite{COHE} we discretize the continuum states of $H_0$ by replacing them with discrete states $|k>$, with eigenvalues $E_k$, separated by a certain energy $\delta$. When $\delta \longrightarrow 0$ one obtains back physical results.} $\left\{|k>\right\}$ of the unperturbed Hamiltonian $H_0$ either directly, or through another discrete eigenvector of $H_0$, the autoionization state, which will be indicated with $|\phi>$.   The autoionization state $|\phi>$ is coupled to $|k>$ through the coupling potential $W$, whose explicit determination will not concern us here and is in fact the subject for a density functional theory approach. We define the coupling strength between $|\phi>$ and $|k>$ through $w_k \equiv <k|W|\phi>$. The coupling between $|\phi>$ and $|k>$ through $W$ modifies the structure of the quasi-continuum states $|k>$. In fact, given the total Hamiltonian $H= H_0 + W$, we can indicate its eigenvalues and eigenvectors respectively with $E_\mu$ and $|\psi_\mu>$, and find the components of $|\psi_\mu>$ on $|\phi>$ and $|k>$ as \cite{COHE}

\begin{eqnarray}
&&<\phi|\psi_\mu> = \frac{1}{\left[1+\sum_{k'}\left(\frac{w}{E_\mu-E_{k'}}\right)^2\right]^{1/2}}\cr &&
<k|\psi_\mu> = \frac{w/(E_\mu-E_k)}{\left[1+\sum_{k'}\left(\frac{w}{E_\mu-E_{k'}}\right)^2\right]^{1/2}}~,
\label{ampli}
\end{eqnarray}
together with the eigenvalue equation

\begin{eqnarray}
\sum_{k}\frac{w^2}{E_\mu-E_{k}}=E_\mu~.
\label{eigen}
\end{eqnarray}
In order to obtain Eq. (\ref{ampli}) and Eq. (\ref{eigen}), we assumed with  \cite{COHE} that all $w_k$ are equal and real, i.e. $w_k=w$, that $<\phi|H_0|\phi> = <\phi|W|\phi>=<k|W|k'>=0$, and that $<k|H_0|k>=E_k = k \delta$. Equivalently, always following \cite{COHE}, we can write Eq. (\ref{ampli}) as

\begin{eqnarray}
&&<\phi|\psi_\mu> = \frac{w}{\left[w^2+ \left(\frac{\hbar \Gamma}{2}\right)^2 + E_\mu^2\right]^{1/2}}\cr &&
<k|\psi_\mu> = \frac{w^2/(E_\mu-E_k)}{\left[w^2+ \left(\frac{\hbar \Gamma}{2}\right)^2 + E_\mu^2\right]^{1/2}}~,
\label{ampli2}
\end{eqnarray}
where the transition rate $\Gamma$ is $\Gamma = 2 \pi w^2/(\hbar \delta)$, and in the limit for $\delta \longrightarrow 0$, $w^2/\delta$ is constant and equal to $\hbar \Gamma / (2 \pi)$.

For each choice of $\mu$ we are left with a two-level system with $|1> = |\psi_0>$ and $|2> = |\psi_\mu>$. The transition amplitude between state $|1>$ and state $|2>$ can be written as

\begin{eqnarray}
&&<2|p|1>  = <\psi_\mu|p|\psi_0> \cr && = <\psi_\mu|\phi><\phi|p|\psi_0>+\sum_k <\psi_\mu|k><k|p|\psi_0>~,
\label{ampli3}
\end{eqnarray}
which can be re-written with the help of Eq. (\ref{ampli2}) as

\begin{eqnarray}
<2|p|1>   = \frac{<\phi|p|\psi_0> w+ \sum_k <k|p|\psi_0> w^2/(E_\mu-E_k)}{\left[w^2+ \left(\frac{\hbar \Gamma}{2}\right)^2 + E_\mu^2\right]^{1/2}}~.
\label{ampli4}
\end{eqnarray}
Finally, the use of the eigenvalue equation Eq. (\ref{eigen}), the definition $v= <\phi|p|\psi_0>$ and the simplifying assumption that $<k|p|\psi_0> = v'$ independently of $|k>$ allows to write

\begin{eqnarray}
<2|p|1>   = \frac{v w+ v' E_\mu}{\left[w^2+ \left(\frac{\hbar \Gamma}{2}\right)^2 + E_\mu^2\right]^{1/2}}~.
\label{ampli5}
\end{eqnarray}
Taking the limit for $\delta\longrightarrow 0$, the quasi-continuum $|\psi_\mu>$ becomes a continuum of states, and with the help of the reduced variables $\mathcal{E} = E_\mu/(\hbar \Gamma/2)$ and $q = \delta v/(\pi w v') = 2 w v/(\hbar \Gamma v')$ we obtain
\begin{eqnarray}
\frac{|<2|p|1>|^2}{v'^2} = \frac{|q+\mathcal{E}|^2}{1+\mathcal{E}^2}~.
\label{fano1}
\end{eqnarray}
In Eq. (\ref{fano1}), $E_\mu$, which is the energy of the continuum state, and can be written as the difference $E_{ph}-E_{12}$, with $E_{12} = \hbar \omega_{12}$ being the a theoretically defined resonance energy, and $E_{ph}$ being the photon energy of the electromagnetic pulse. By this, $E_{\mu}$ can be positive or negative. This is consistent with previous choice to take the discrete energy level $|\phi>$ as the origin of the energy axis, that is $<\phi|H_0|\phi>=0$.

We reproduced the well-known result that the probability of exciting the atom from the ground state follows a Fano profile as a function of $E_{ph}$.  As said before, the cross-section is proportional to the square of the transition amplitude, and can therefore be written as

\begin{eqnarray}
\sigma = \sigma_b \frac{(q+\mathcal{E})^2}{1+\mathcal{E}^2}~,
\label{sigsig3}
\end{eqnarray}
which corresponds to Eq. (\ref{sigsig2}) with $\sigma_a = 0$. A generalization to the case with $\sigma_a$ non zero can be obtained considering that transitions to the continuum may not interact with the autoionization state, as discussed before. This can be readily done by introducing a the background cross-section $\sigma_a$, and thus recovering Eq. (\ref{sigsig2}).

\section{\label{seven} Semi classical treatment of light propagation through a cell
containing resonantly absorbing gas}

In this article we model the interaction of the gas in the cell with the FEL radiation response as the response of a macroscopic medium under the action of classical fields. Neglecting magnetic interaction and non-linear effects, the electric field $\vec{E}$ of the incident pulse induces a macroscopic polarization vector $\vec{P}$ in the medium, which is linearly related to $\vec{E}$ through the susceptibility $\chi$ according to the well known relation

\begin{eqnarray}
\vec{P} = \epsilon_0 \chi \vec{E}~.
\label{PP}
\end{eqnarray}
The susceptibility $\chi$ is a function of the the frequency of the applied field, and is related to the complex refractive index $n$ via

\begin{eqnarray}
n^2 = 1 + \chi~,
\label{nn}
\end{eqnarray}
and to the dielectric constant through

\begin{eqnarray}
\epsilon = \epsilon_0(1 + \chi)~.
\label{dc}
\end{eqnarray}
The determination of the susceptibility $\chi$ depends on the dynamics of the system. For example, as is well-known, if the perturbation excites transitions from the ground state $|1>$ to an excited state $|2>$ of the atoms in the medium, one finds quantum-mechanically the usual result in agreement with the damped oscillator model:

\begin{eqnarray}
\chi(\omega) = \frac{N |<2|p|1>|^2}{3 \epsilon_0 \hbar V} \left(\frac{1}{\omega_{12} - \omega - i\gamma}\right)
\label{oscmod}
\end{eqnarray}
where $N$ is the number of atoms in the volume $V$, the operator $p$ has been defined in Eq. (\ref{P}), $\omega_{12}$ is the transition frequency, and $\gamma$ models the radiative loss rate. Eq. (\ref{oscmod}) obeys, as well-known, a complex Lorentian relation.

In principle one may calculate an analogous of Eq. (\ref{oscmod}) in the case when $|2>$ is a state in the continuum, coupled with atom in a doubly excited state, that is our autoionization state. In this case there are two contributions to the susceptibility. One arises from the dipole moment between the two states of the atom, and the other from the response to direct photo effect, that is the electromagnetic transition  from ground state to continuum. In this article we do not present this direct calculation, and we make use of general properties of the susceptibility instead.

As is also well known, the susceptibility $\chi$, and as a consequence the refractive index and the dielectric constant, obey Kramers-Kroning relations, and it is always possible to recover the imaginary part of one of these quantities from the a-priori knowledge of the real part, or viceversa. The typical example is the relation between real and imaginary part of Eq. (\ref{oscmod}). In our case, the a-priori information available is the knowledge of the cross-section, which obeys a Fano profile. The cross-section and the imaginary part of the refractive index are related through:

\begin{eqnarray}
\mathrm{Im}[k n] = \frac{n_0}{2} \sigma
\label{crossn}
\end{eqnarray}
where $n_0$ is the gas density and $k=\omega/c$. As a result, as explained in Section \ref{TTT}, the prolongation of the logarithm of the transmittance to the upper complex plane is analytical, and the phase of the transmittance can be recovered from the knowledge of the cross-section. Once the transmittance $T(\omega)$ is known, the product with the incident field $E(\omega)$ yields the transmitted pulse $E(\omega) T(\omega)$, which can then be characterized in the time domain by means of a Fourier transform.

\subsection{Collision line broadening}

As seen before, the knowledge of the cross-section allows for a full characterization of the medium behavior in the frequency domain in terms of transmittance. This is enough for our purposes, since the transmitted pulse can be recovered in the time domain by applying a Fourier transform. Similarly, the time-domain evolution of the macroscopic dipole moment $P(t)$ can be found by Fourier transforming Eq. (\ref{PP}), which is based on the knowledge of $\chi(\omega)$, recovered through the cross-section, and of the incident pulse in the frequency domain. The macroscopic dipole moment $P(t)$ decays in a time scale of the order of $\gamma^{-1} \sim 100$ fs, the inverse of the ionization rate.

In this paragraph we will consider collision line broadening as a possible decay mechanism, and show that in our case of interest it can be neglected. Following, for example, \cite{WIEN}, if the mean free path between two collision is much longer than the gas cell dimensions, one can consider the atoms in the gas moving along straight-line trajectories, colliding in a time which is much shorter than the radiative lifetime of the atomic excited state. Moreover, the time between different collision is long compared with such lifetime too. Then, the collision can be modeled as a loss of coherence due to a sudden phase jump of the atomic excited wavefunction. The time $\tau_{\mathrm{col}}$ between two collisions, i.e. the time between sudden phase jumps, is the inverse of the collisional rate $\gamma_\mathrm{col}^{-1}$.  Always following \cite{WIEN} we demonstrate that collision line broadening can be neglected in our case. In order to do so, we first observe that, for hard-sphere collisions between atoms of mass $M$ and radius $R_a$ and assuming a gas density $n_0$, the inverse of the collisional rate is

\begin{eqnarray}
\gamma_\mathrm{col}^{-1} = \tau_{\mathrm{col}} \simeq \frac{\sqrt{M / (\pi kT)}}{8R_a^2 n_0}
\label{taucol}
\end{eqnarray}

In our case for the Helium gas we have $R_a  =  1.2 {\AA}$,  $n_0 = 3 \cdot 10^{16} cm^{-3}$  (for
a pressure of $1$ Torr at room temperature) $M c^2 \simeq 4 10^9$ eV,  $kT \simeq  0.03$ eV, and as a result we obtain  $\tau_{\mathrm{col}}  \simeq  400$ ns. Since this time scale is much longer than any other time scale in our system, we can neglect this effect.
In other words, the decay of the macroscopic polarization  is due to spontaneous radiation only.

\section{Operation of the LCLS-II soft X-ray FEL by use of the self-seeding scheme with gas monochromator. A feasibility study.}

\begin{figure}[tb]
\includegraphics[width=1.0\textwidth]{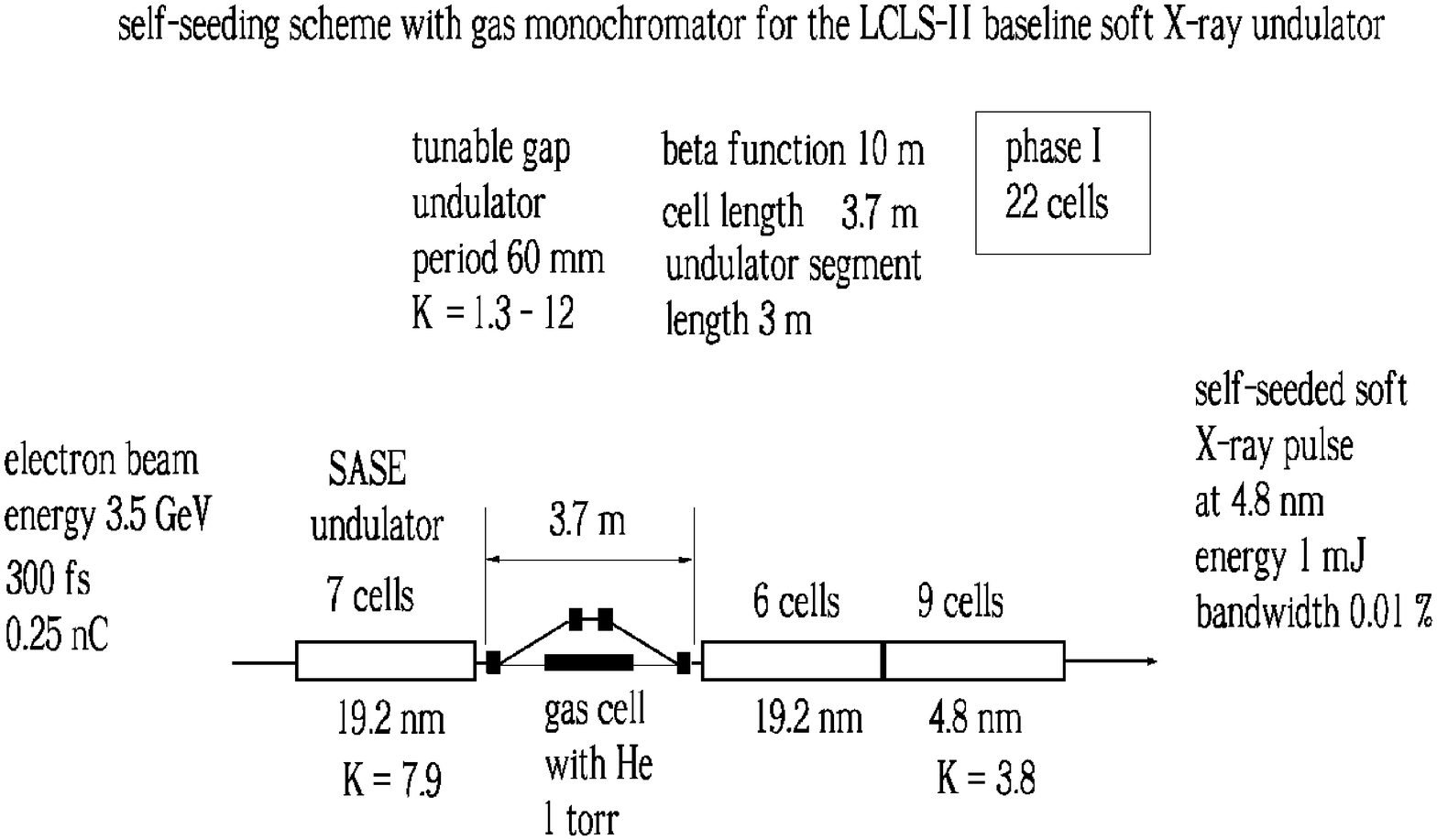}
\caption{Design of the LCLS-II  baseline undulator system for generating of generating of highly monochromatic, high power soft X-ray pulses. The method exploits a combination of a self-seeding scheme with gas monochromator and gap-tunable undulator for harmonic generation.} \label{SXR1}
\end{figure}

\begin{figure}[tb]
\includegraphics[width=1.0\textwidth]{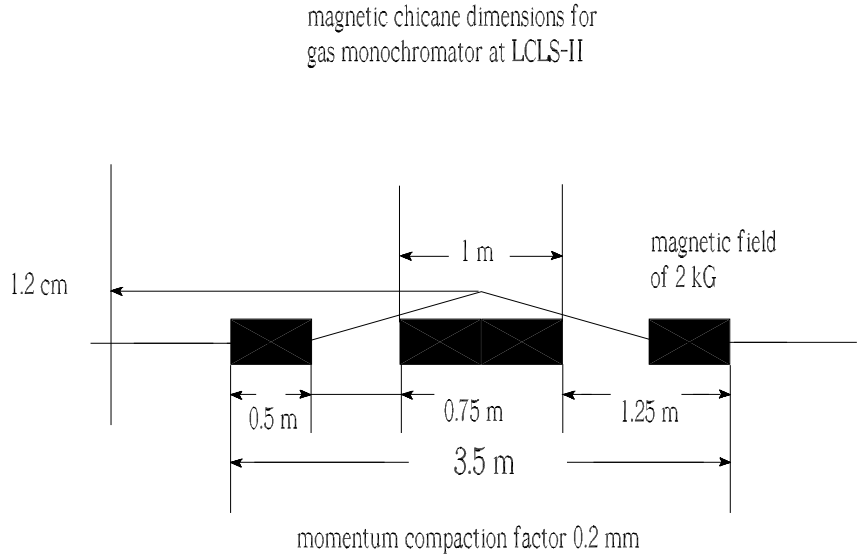}
\caption{Magnetic chicane dimensions for gas monochromator installation at
LCLS-II} \label{SXR2}
\end{figure}

\begin{figure}[tb]
\includegraphics[width=0.5\textwidth]{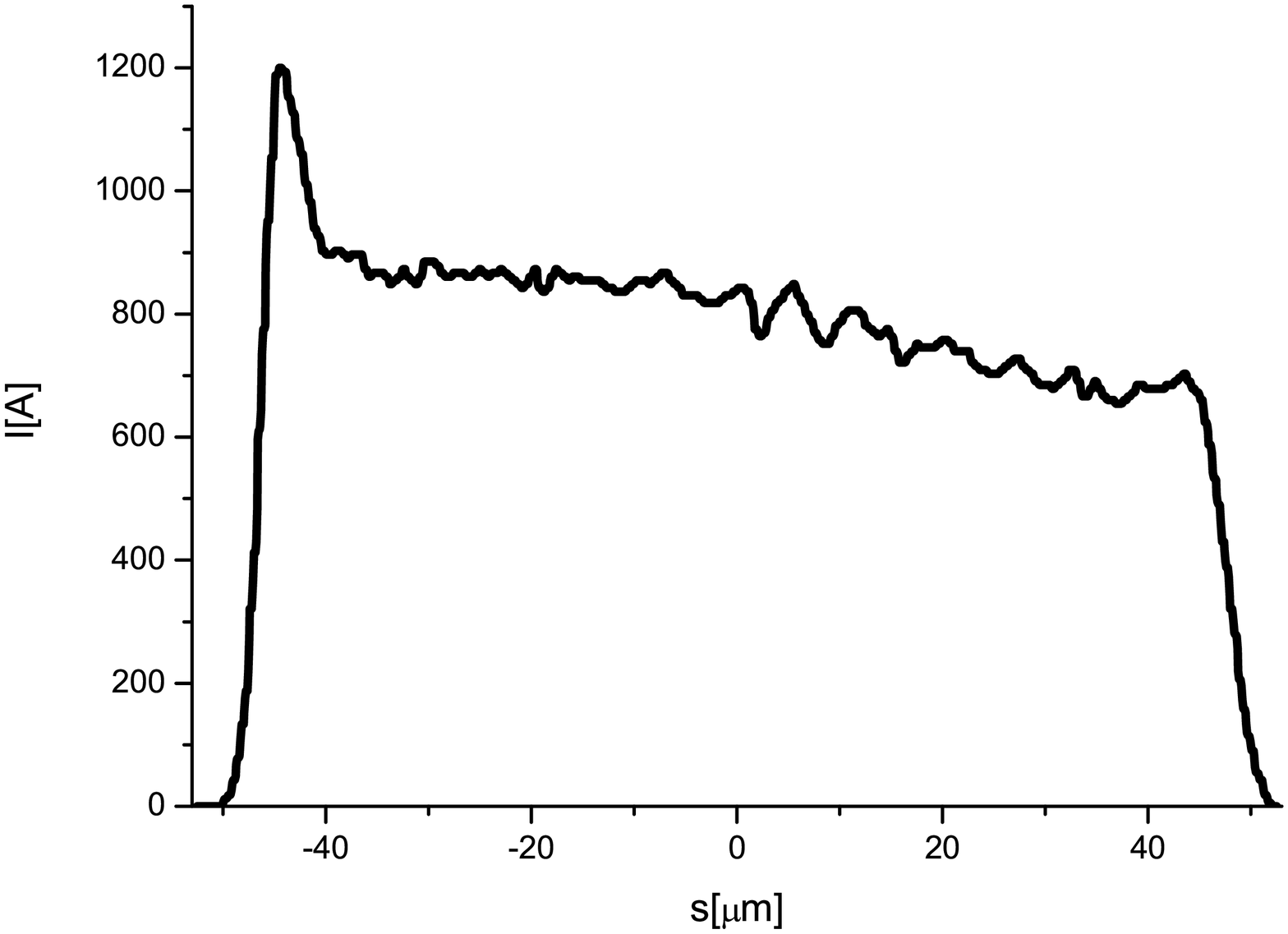}
\includegraphics[width=0.5\textwidth]{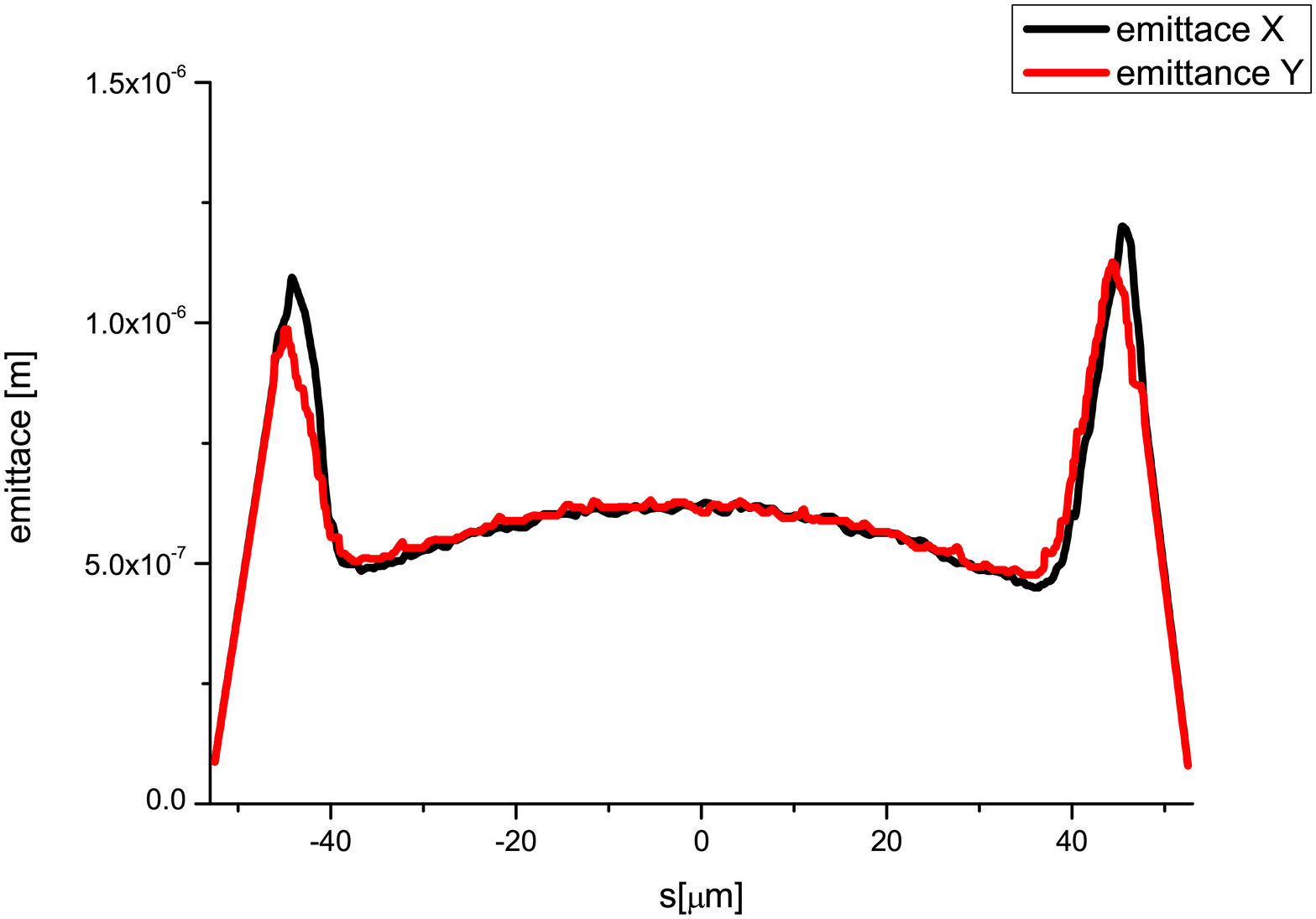}
\includegraphics[width=0.5\textwidth]{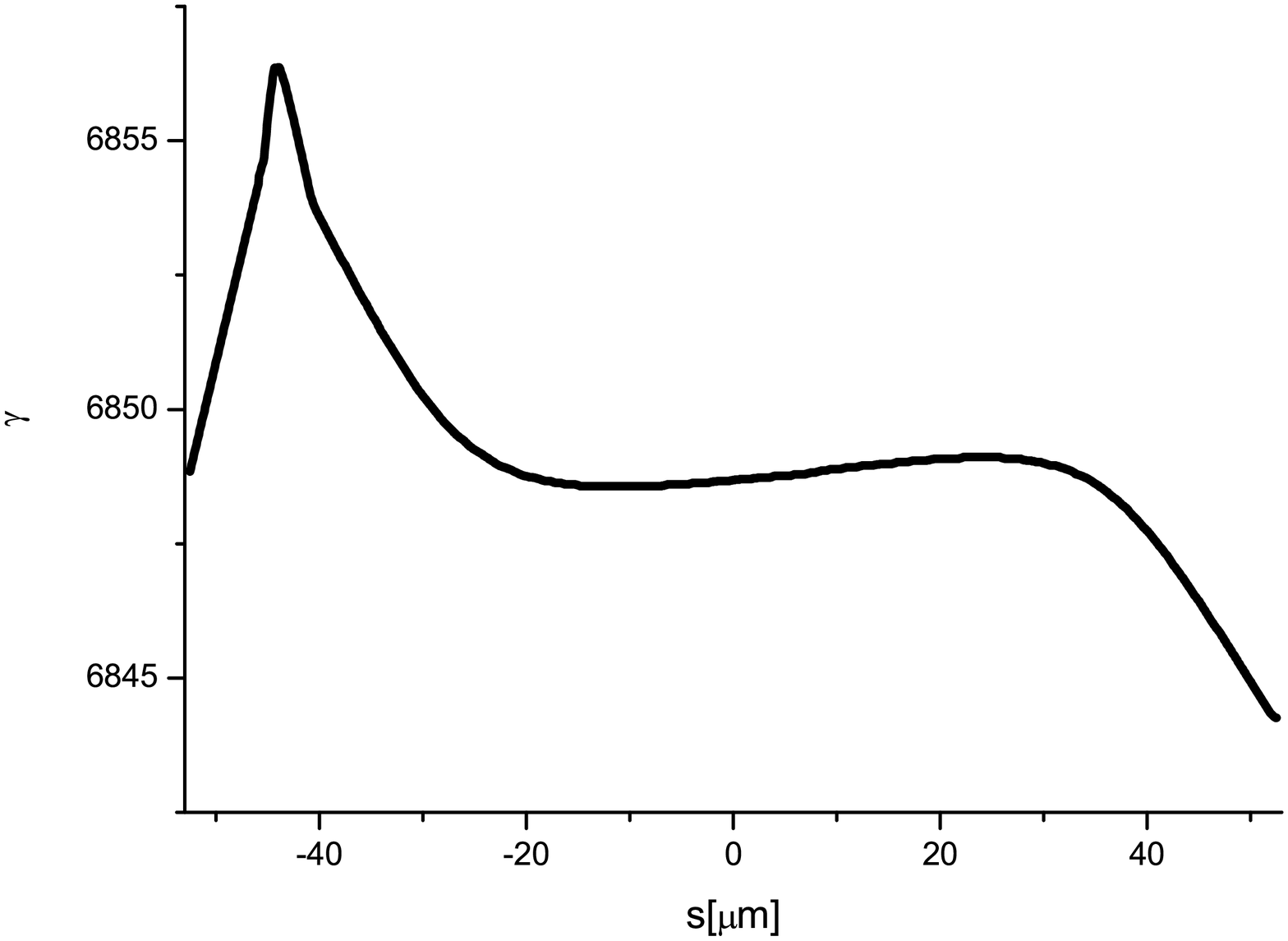}
\includegraphics[width=0.5\textwidth]{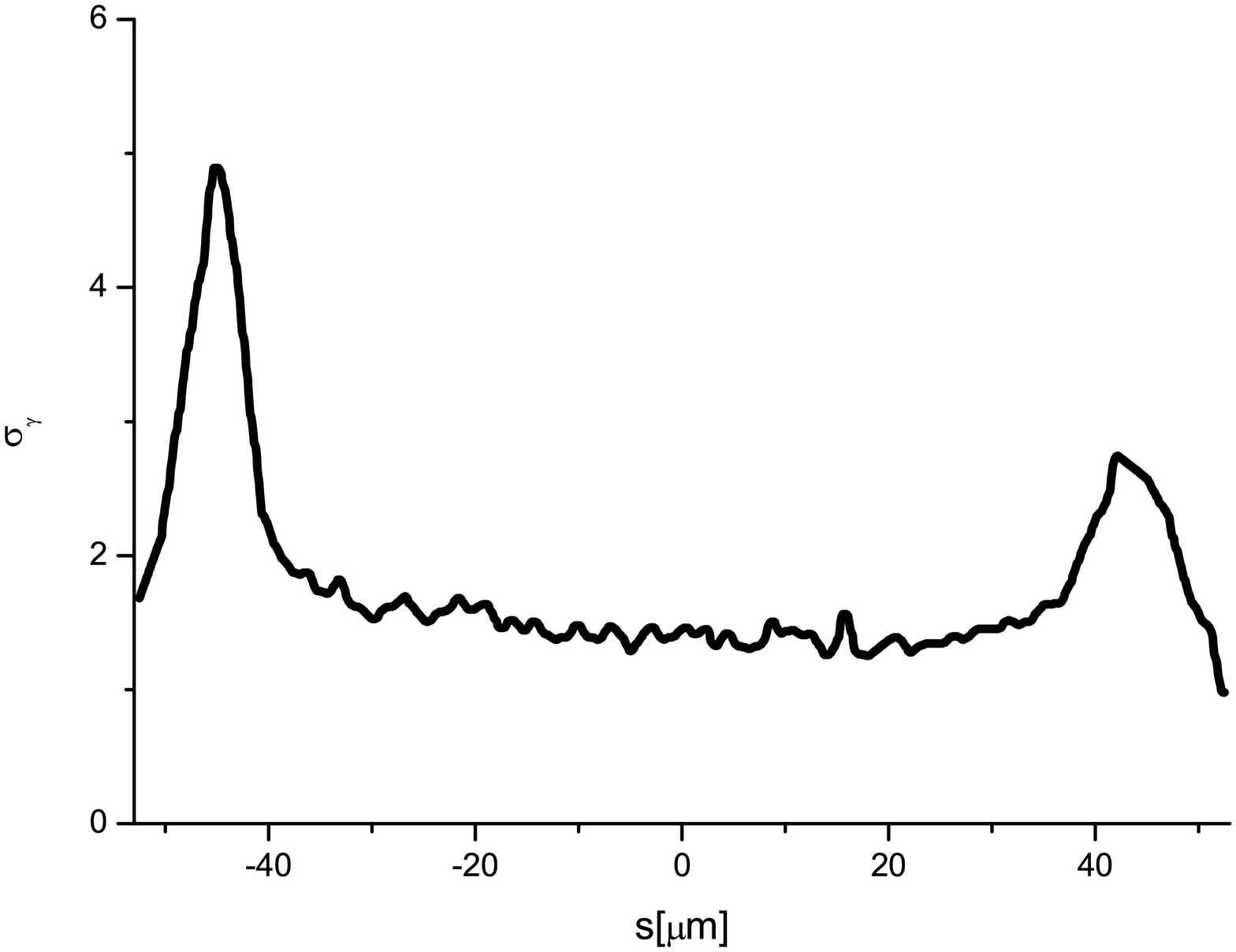}
\caption{Electron beam characteristics at the entrance of the setup at the LCLS. (upper left) Current profile. (upper right) horizontal and vertical geometrical emittance. (lower left) energy profile. (lower right) rms energy spread profile.} \label{ebeam}
\end{figure}

The autoionizing resonance that we consider here as an example is in the $20$ nm-range. In order to generate shorter wavelengths we need to go to higher harmonic numbers. A possible way of doing so is to use a two-stage output undulator, with the second stage resonant to one of the harmonics of the first one, Fig. \ref{SXR1}. The magnetic chicane dimensions are sketched in Fig. \ref{SXR2}. In our feasibility study we will consider, for exemplification purposes, the LCLS-II setup equipped with a 80 m long gap-tunable undulator, which can be operated within the photon energy range between $30$ eV to $1100$ eV at 3.5 GeV electron beam energy.  Due to specific of Front End Enclosure  (FEE) system of LCLS-II will be possible direct the radiation into experimental stations only starting around the 200 eV photon energy range. With the chosen parameters \footnote{We used a not updated set for the LCLS-II SXR undulator parameters for our simulations, but the same wavelength is achievable for the current design too.} (see Table \ref{tt2} and Fig. \ref{ebeam}), the amplitude of the 4th harmonic of the density modulation at the exit of first stage of the output undulator is high enough and dominates significantly over the amplitude of shot noise harmonic. This modulation density serves as an input signal for the second part of output undulator, which is indeed resonant with the 4th harmonic. An important feature of our design is that no dispersion section is introduced between the two stages in the output undulator. The advantage of this method of harmonic generation in self-seeded FELs is in the simple hardware required (only a short magnetic chicane for the gas cell installation). However, a gap-tunable baseline undulator is needed.

\begin{table}
\caption{Parameters for the nominal pulse mode of operation used in
this paper.}

\begin{small}\begin{tabular}{ l c c}
\hline & ~ Units &  ~ \\ \hline
Undulator period      & mm                  & 60     \\
K parameter      & -                   & 1.3-12  \\
Wavelength  (fundamental)               & nm                  & 19.2   \\
Energy                & GeV                 & 3.5   \\
Charge                & nC                  & 0.25 \\
\hline
\end{tabular}\end{small}
\label{tt2}
\end{table}

Here we focus on the implementation of our self-seeding scheme exploiting a wake monochromator to generate fully coherent soft X-rays for LCLS-II. The tentative design for the gas monochromator self-seeding  technique at the LCLS-II aims at  generating  fully coherent radiation at about $5$ nm.
The main parameters  for the nominal 5 nm case are listed in Table \ref{tt2} in the next Section.  Full 3D have been performed in order to confirm the scheme feasibility. The simulations is performed with the code GENESIS 1.3 code \cite{GENE}, which uses as input the beam parameters obtained in start to end simulations \cite{echo}.    The beam distributions are shown in Fig. \ref{ebeam} showing the current profile, the horizontal and vertical geometrical emittances, the energy profile and the rms energy spread profile. The average betatron function used was $\beta = 10$ m.

In all simulations we assumed that the gas used in the monochromator is Helium, and that the energy of the VUV FEL photons is $64.4$  eV, or $19.2$ nm, corresponding to the $n=4$ autoionization profiles for the  $(sp,2n+)$ autoionizing series of Helium, discussed in Section \ref{phab}.

\begin{figure}[tb]
\includegraphics[width=0.5\textwidth]{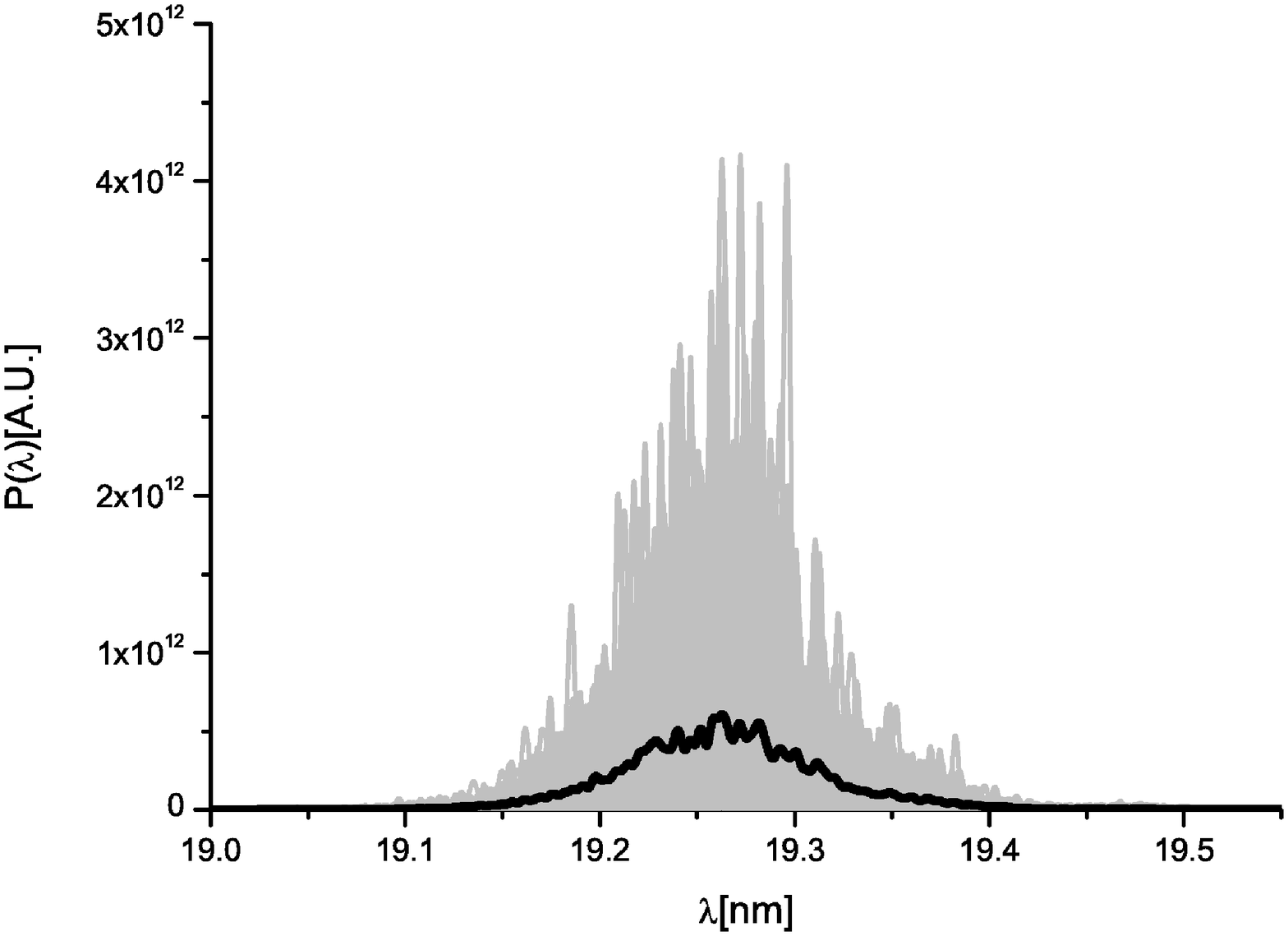}
\includegraphics[width=0.5\textwidth]{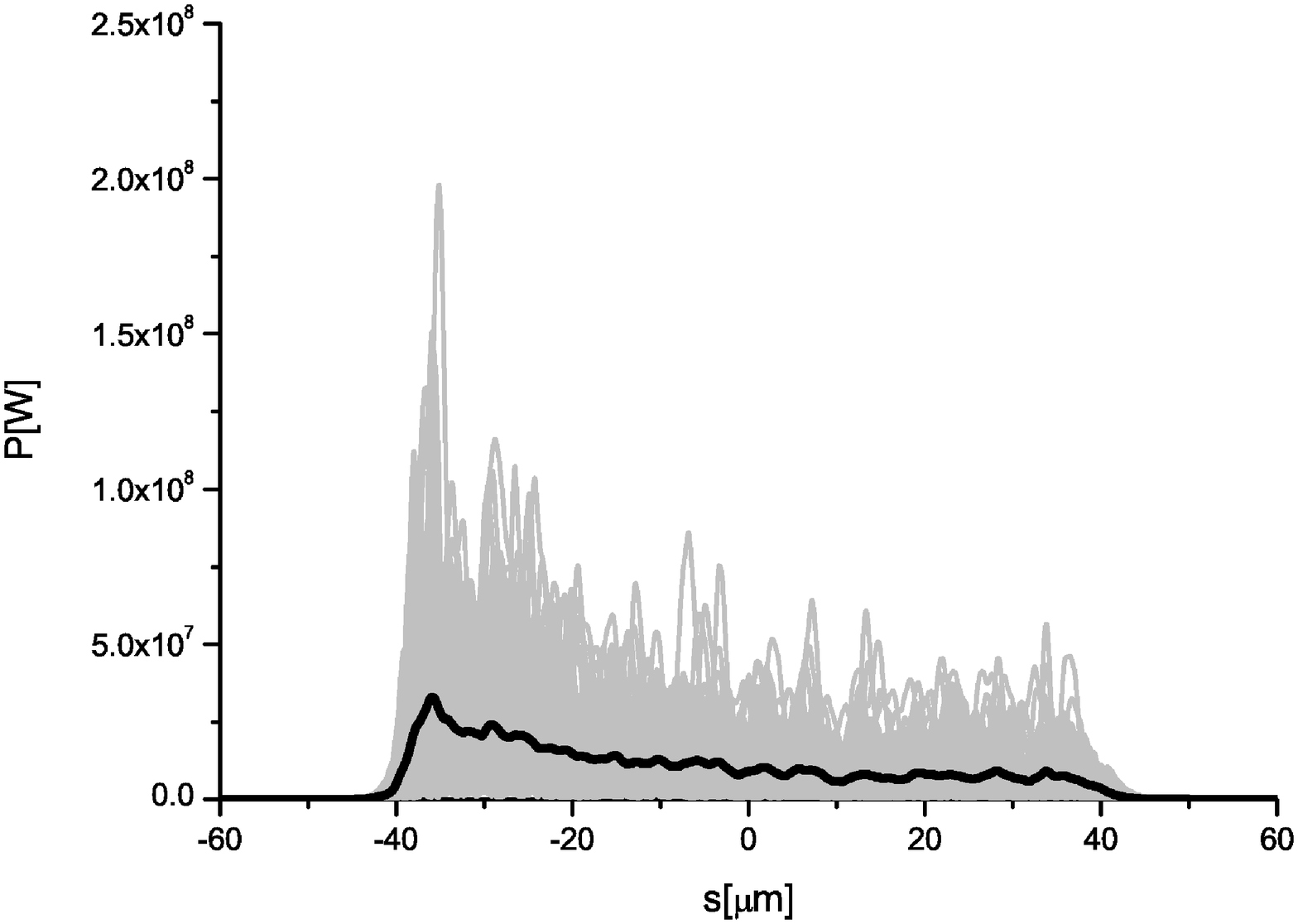}
\caption{Pulse spectrum (left plot) and power (right plot) before the gas cell. Grey lines refer to single shot realizations, the black line refers to an average over one hundred realizations.} \label{SXR2S}
\end{figure}

\begin{figure}[tb]
\includegraphics[width=0.5\textwidth]{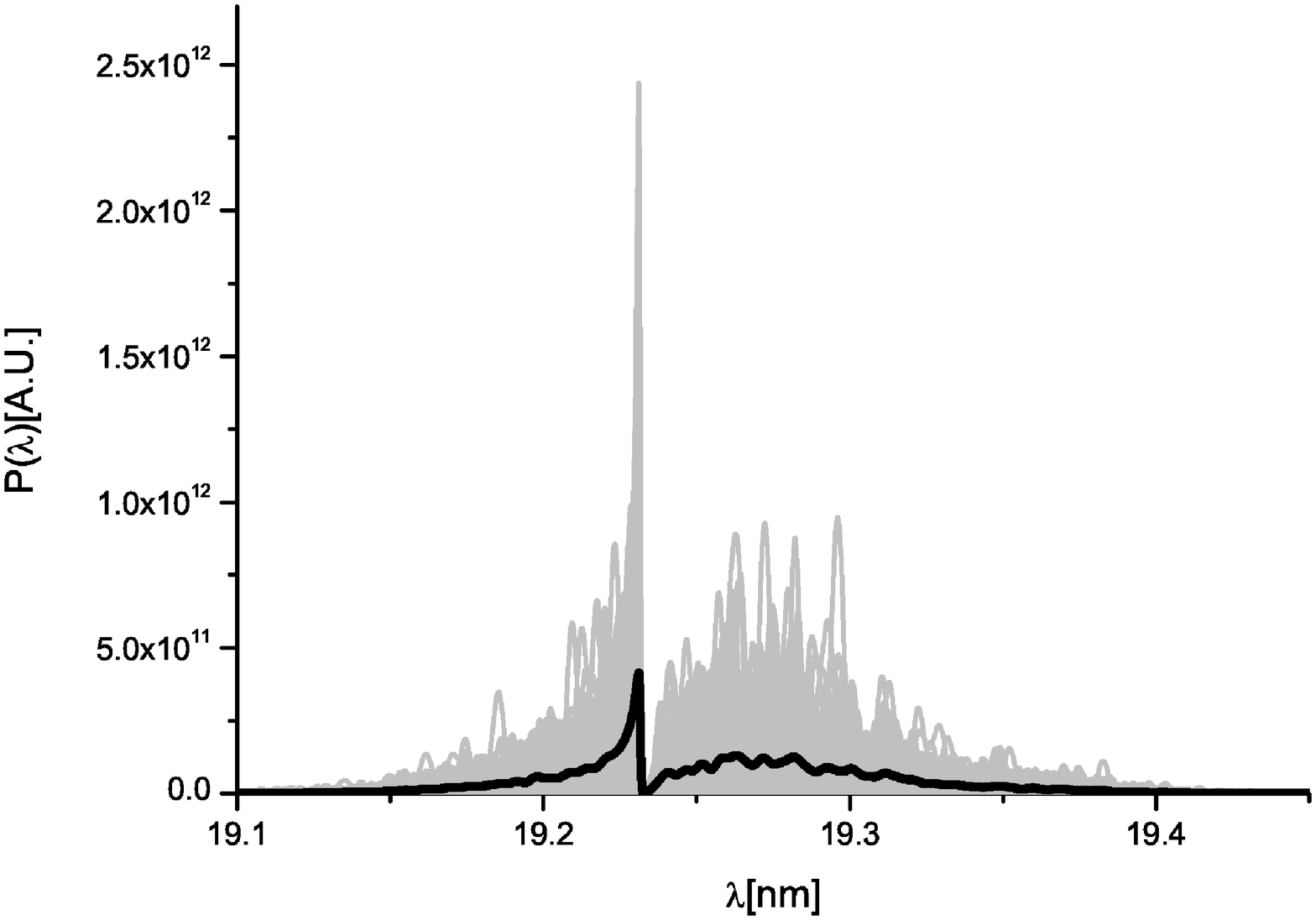}
\includegraphics[width=0.5\textwidth]{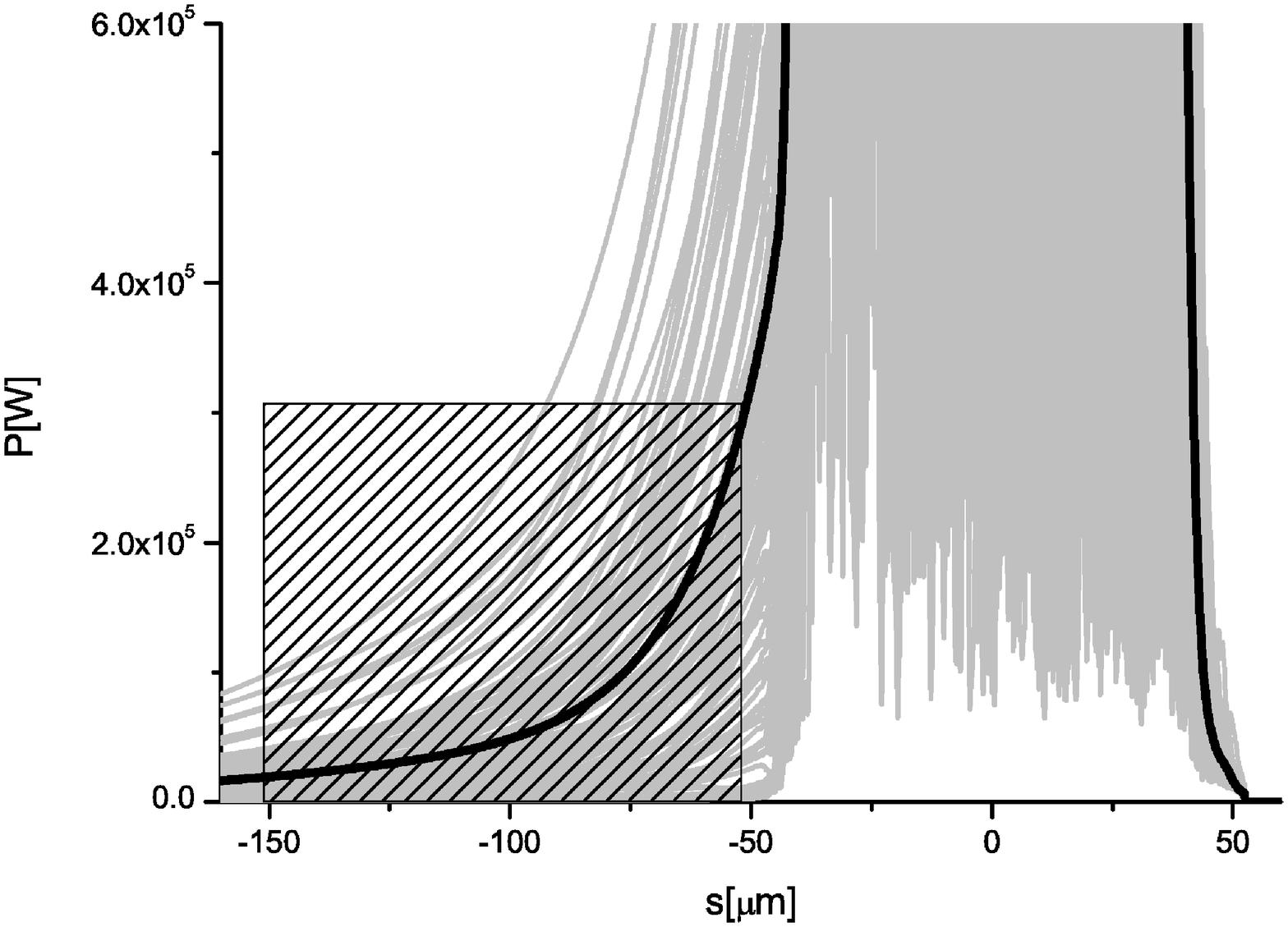}
\caption{Pulse spectrum after the gas cell (left plot) and seed power (right plot). Grey lines refer to single shot realizations, the black line refers to an average over one hundred realizations.} \label{SXR3S}
\end{figure}

\begin{figure}[tb]
\includegraphics[width=1.0\textwidth]{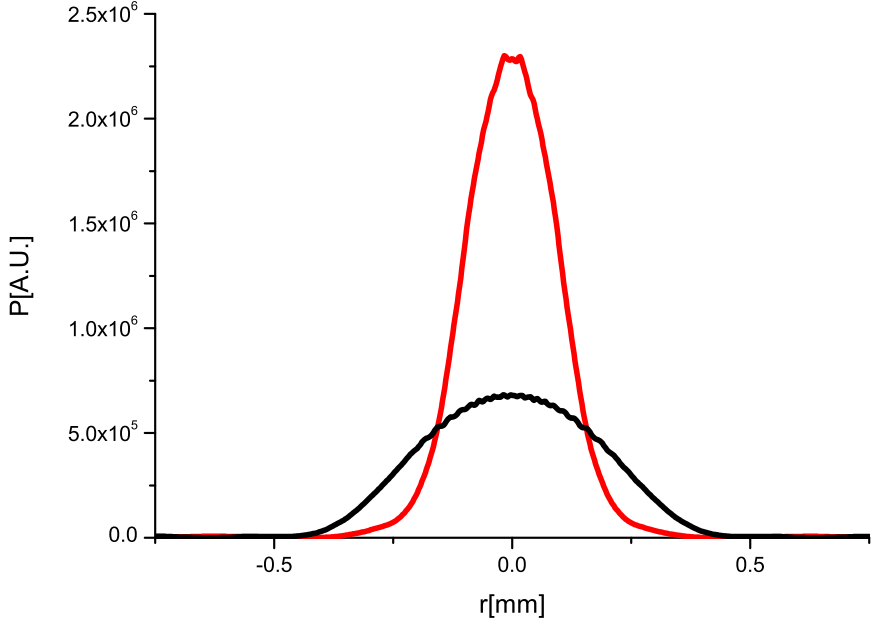}
\caption{Output size of the FEL radiation pulse. The black line is the photon beam profile before the monochromator, the red line refers to the photon beam profile after $4$ m propagation, that is at the exit of the differential pumping system.} \label{size}
\end{figure}

The spatial and spectral structure of the LCLS-II pulse at $19.2$ nm wavelength and at the location of the gas monochromator are shown\footnote{From Fig. \ref{SXR2S} and Fig. \ref{size} we can estimate a photon density of $10^{15}$ cm$^{-2}$. It should be noted that the resonance line is about two orders of magnitude narrower ($10^{-4}$) compared with the SASE bandwidth ($5\cdot 10^{-3}$), while the maximum cross section reaches about $10$ Mb. As a result, attenuation takes place mainly due to the background cross-section, of about $1.5$ Mb. Recalling that $1$ Mb is $10^{-18}$ cm$^{2}$ one can confirm the validity of the perturbation theory approach used to calculate the overall transmission function of the gas cell.} in Fig. \ref{SXR2S}. The effect of the passage of the photon pulse through the gas cell is shown in Fig. \ref{SXR3S}, where one can clearly see the bandstop filtering in terms of spectrum, and its counterpart in the time domain in terms od seeding pulse. The FEL radiation pulse has a transverse FWHM size of about $0.5$ mm  at the exit of the differential pumping system, the red line in Fig. \ref{size}.

While the photon beam passes through the gas cell, the electron crosses the chicane, where the microbunching is removed.  In fact, for Gaussian local
energy spread, the amplitude of the density modulation $a$ at the chicane exit is given by

\begin{eqnarray}
a = a_0 \exp\left[-\frac{1}{2}\frac{\left\langle(\delta
\gamma)^2\right\rangle}{\gamma^2}\frac{R_{56}^2
}{\lambdabar^2}\right] \label{uno}
\end{eqnarray}
where $a_0$ is the amplitude of the density modulation at the entrance of the chicane, $R_{56}$ is the momentum compaction factor and $\lambdabar$ is the reduced wavelength. In our case, parameters of interest are the dispersion $R_{56} = L_w \theta^2 \sim 0.2$ mm (insuring a delay equal to $0.1$ mm), and the relative energy spread $\Delta \gamma/\gamma \sim 0.03 \%$, corresponding to an energy spread of about $1.5$ MeV at an energy of $3.5$ GeV. For a wavelength of $19$ nm, these parameters lead to the suppression of the beam modulation by a factor of about $\exp(-200)$.

\begin{figure}[tb]
\includegraphics[width=0.5\textwidth]{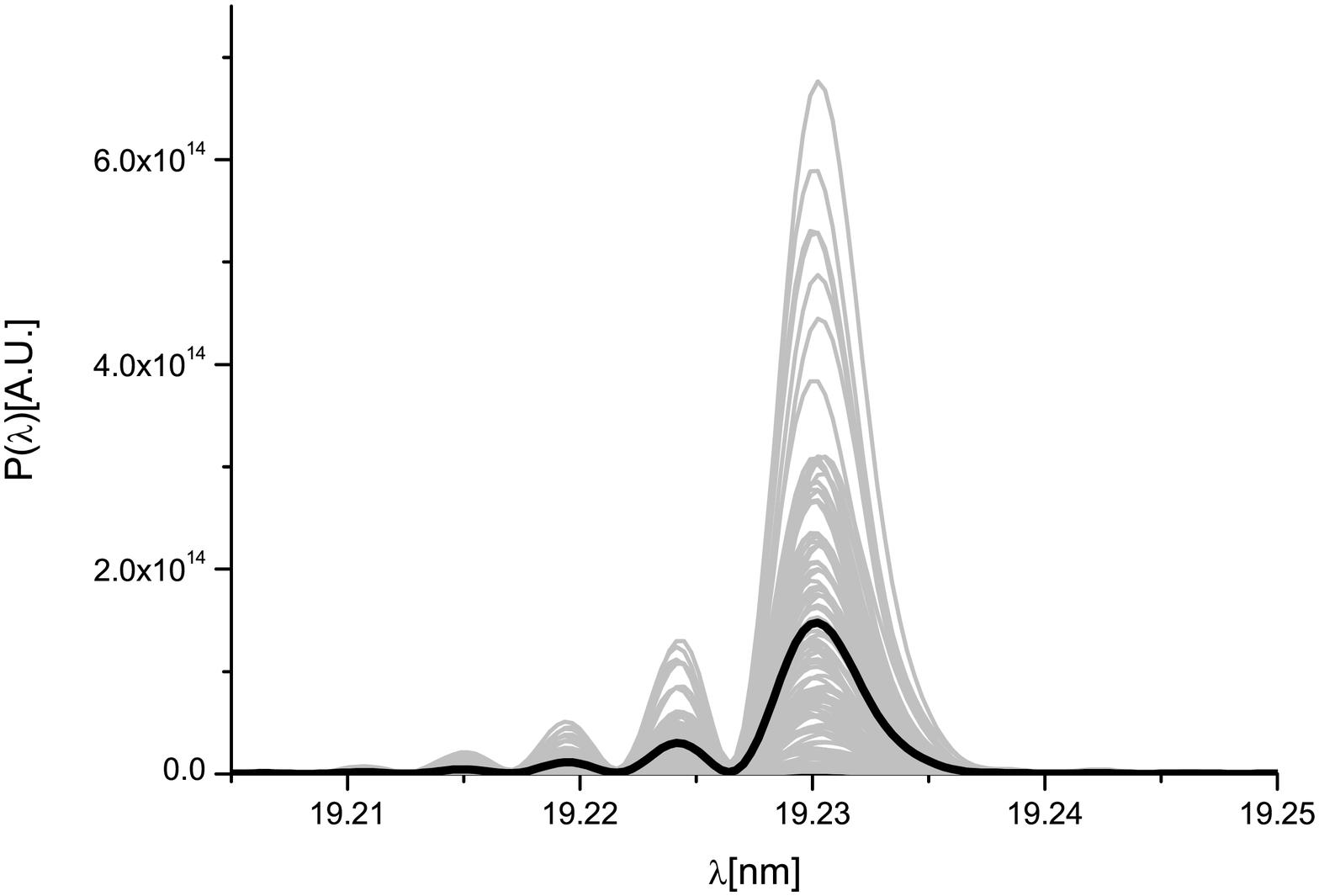}
\includegraphics[width=0.5\textwidth]{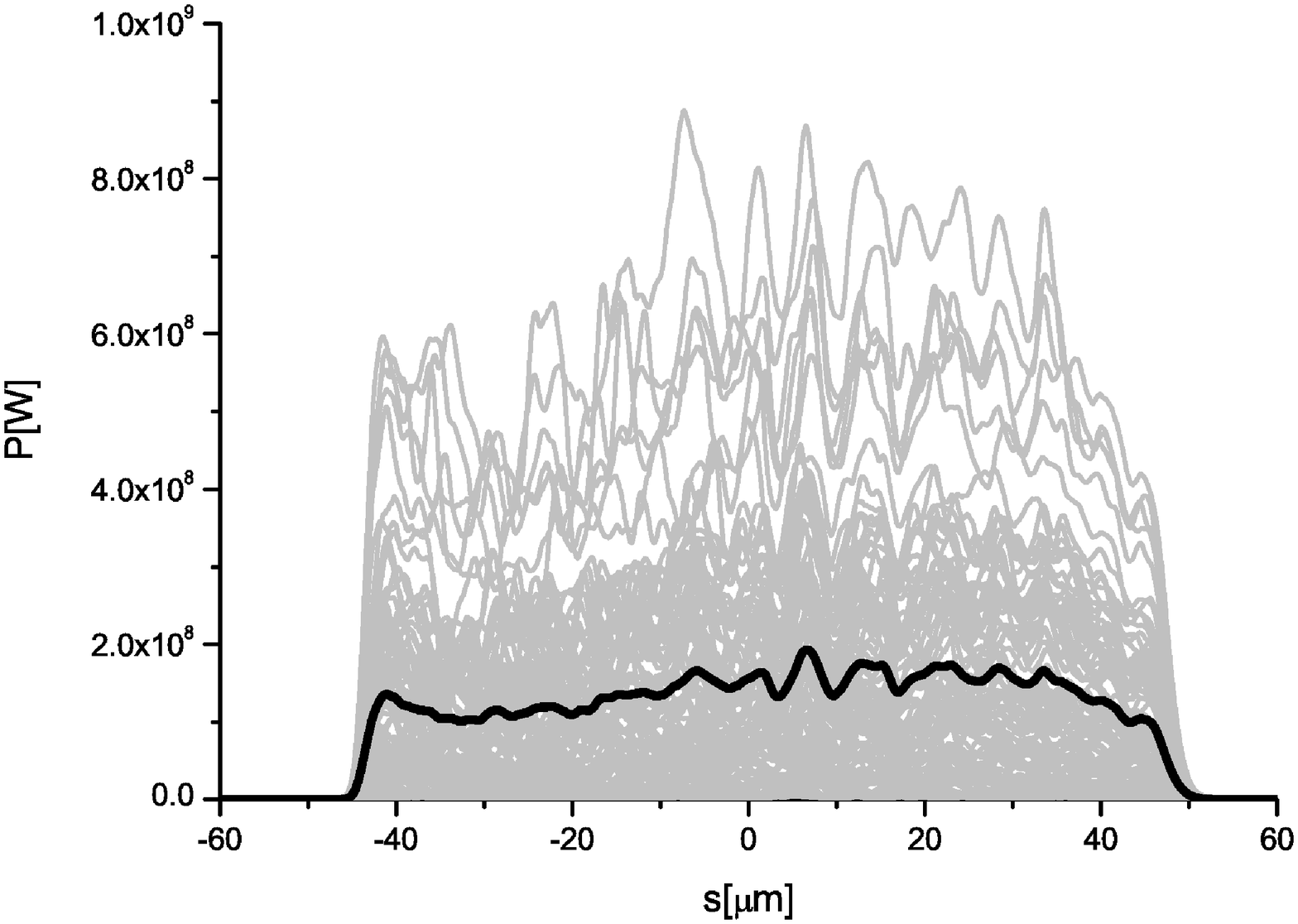}
\caption{Output characteristics for the second undulator tuned at the fundamental harmonic. Spectrum (left plot) and power (right plot). Grey lines refer to single shot realizations, the black line refers to an average over one hundred realizations.} \label{SXR4b}
\end{figure}

Following the chicane, the electron beam is recombined with the photon beam at the entrance of the output undulator. The first stage of the output undulator, consisting of the first $6$ cells, is tuned at the fundamental, while the second stage, consisting of $9$ cells, is tuned at the fourth harmonic, Fig. \ref{SXR1}. The output characteristics in terms of power and spectra at the exit of the first stage are shown in Fig. \ref{SXR4b}.

\begin{figure}[tb]
\includegraphics[width=0.5\textwidth]{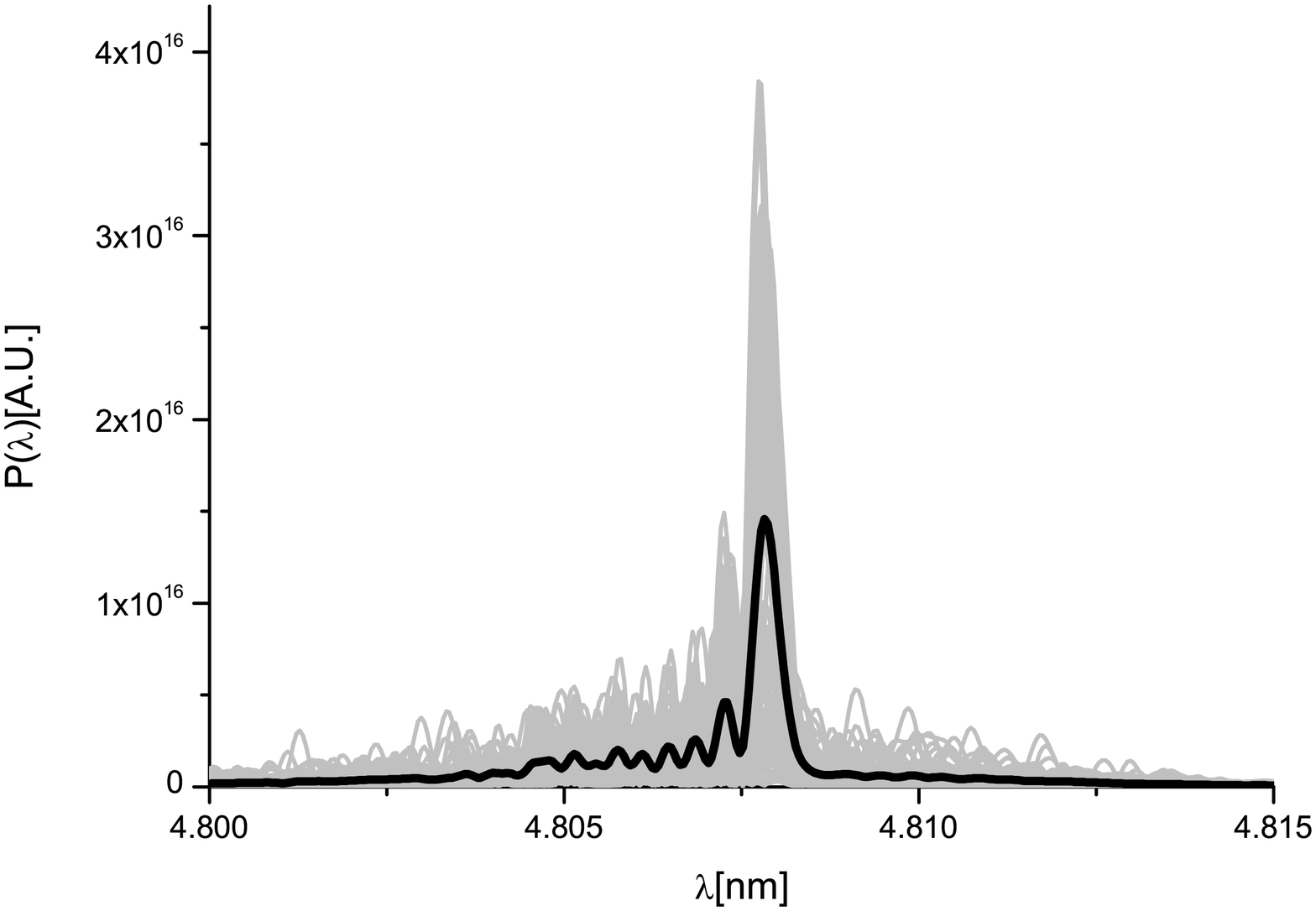}
\includegraphics[width=0.5\textwidth]{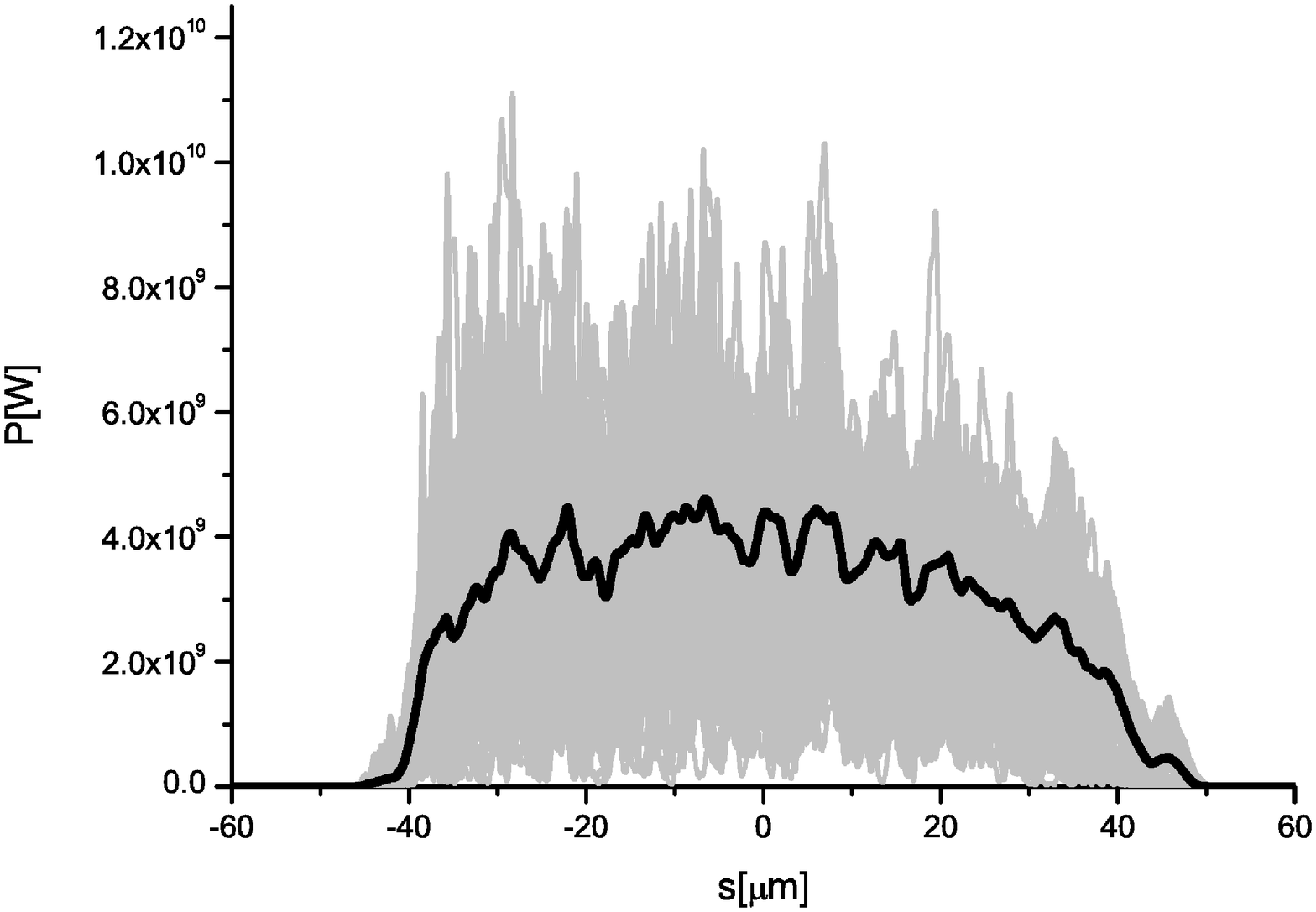}
\caption{Output characteristics for the second undulator tuned at the fourth harmonic. Spectrum (left plot) and power (right plot). Grey lines refer to single shot realizations, the black line refers to an average over one hundred realizations.} \label{SXR5b}
\end{figure}

\begin{figure}[tb]
\includegraphics[width=0.5\textwidth]{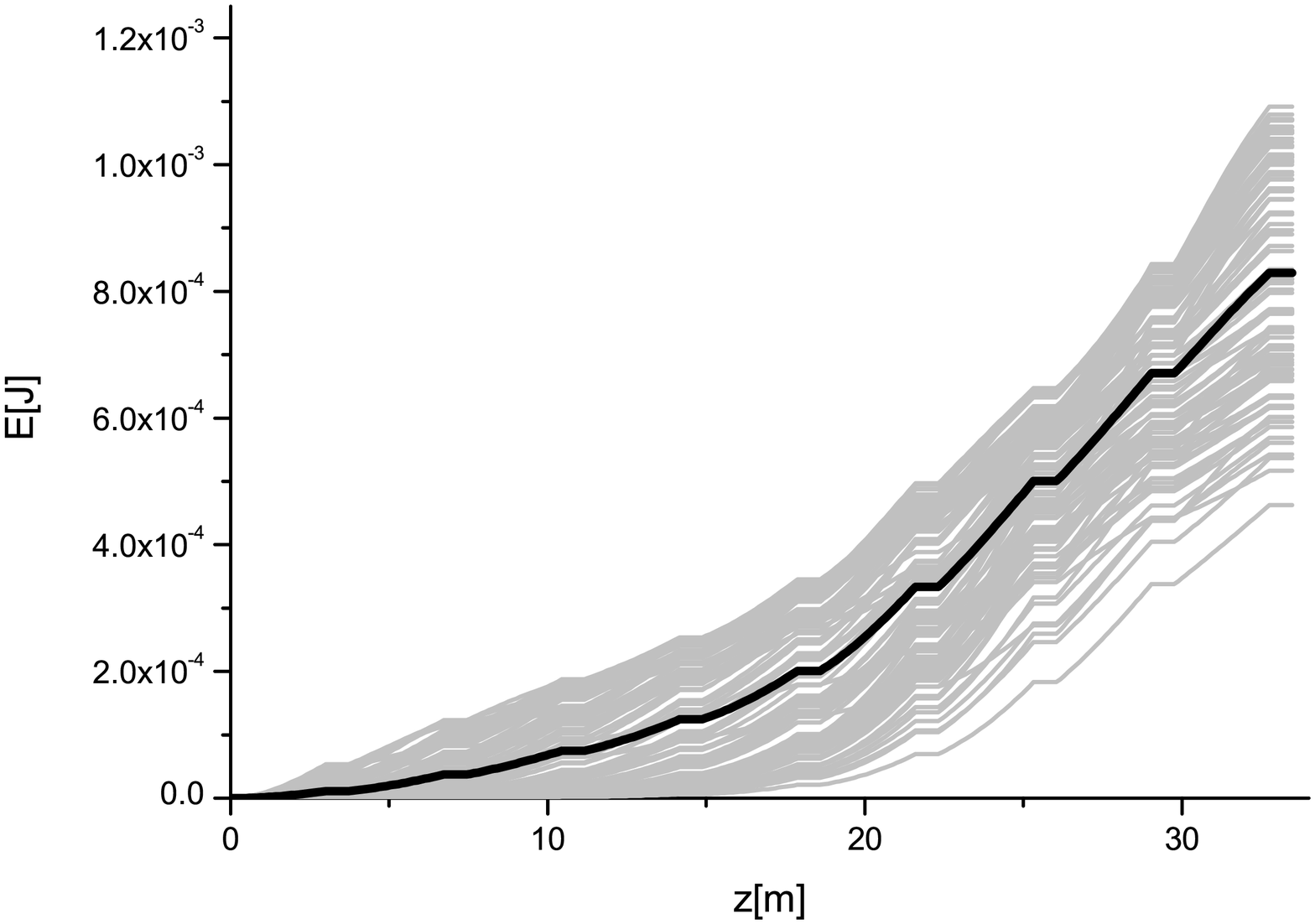}
\includegraphics[width=0.5\textwidth]{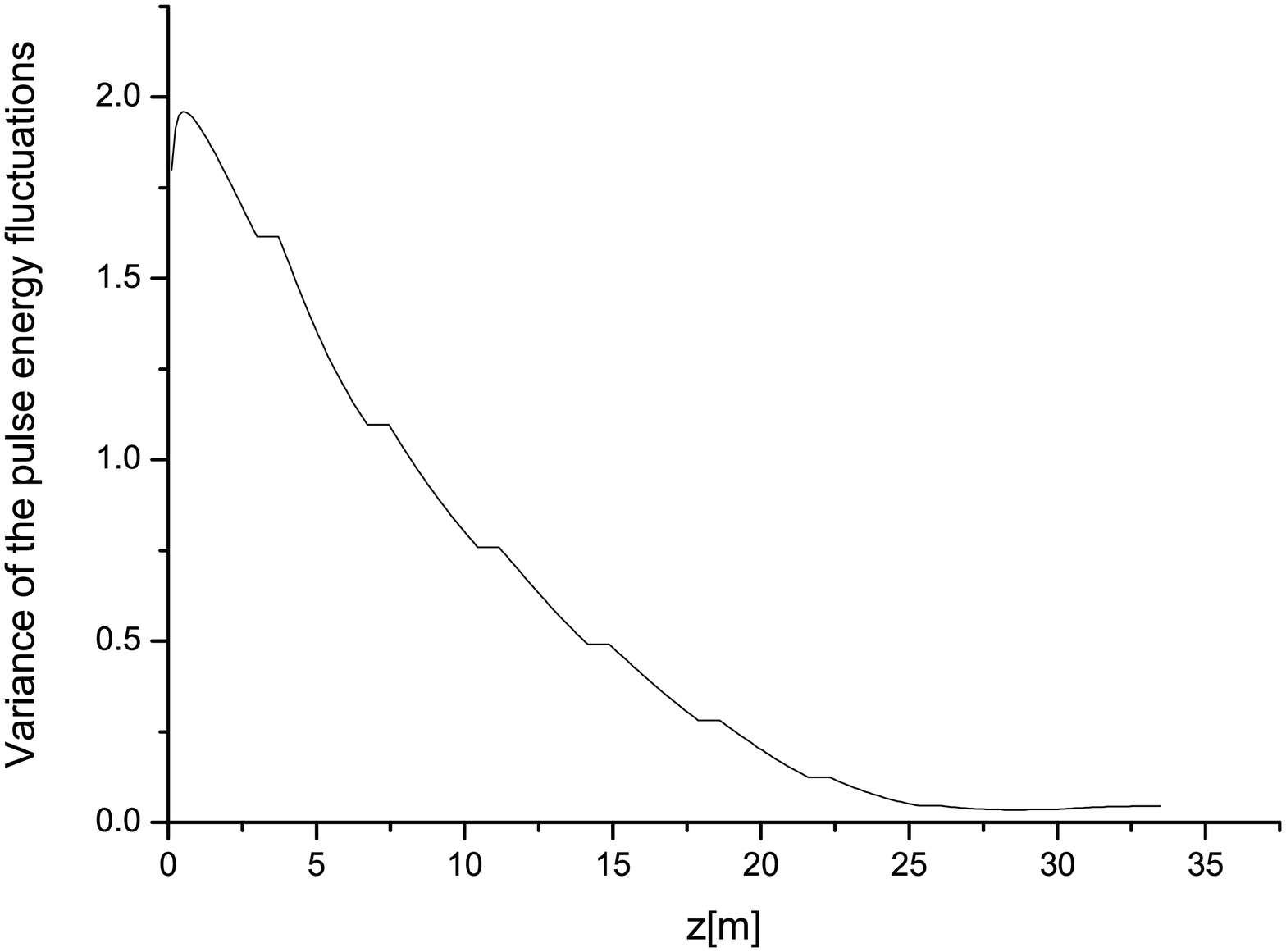}
\caption{Output characteristics for the second undulator tuned at the fourth harmonic, as a function of the length of the second undulator. Energy (left plot) and variance of the energy fluctuations (right plot). Grey lines on the left plot refer to single shot realizations, the black line refers to an average over one hundred realizations.} \label{SXR6b}
\end{figure}
The final output at the fourth harmonic is shown in Fig. \ref{SXR5b}, showing power and spectrum, and in Fig. \ref{SXR6b}, showing energy and variance of the electron energy fluctuations. A mJ, fully coherent pulse with a bandwidth in the order of the $0.01 \%$  is produced at a wavelength of $4.8$ nm.

\section{Conclusions}

In this paper we propose a method for controlling the line width of VUV and soft X-ray SASE FELs that offers simplicity and flexibility, and can be added to the baseline undulators of many facilities without significant cost or design changes. Monochromatization down to the Fourier transform limit of the radiation pulse can be performed by exploiting an almost trivial setup composed of as few as two components. The key components of such scheme
include only a cell containing dilute noble gas, and short magnetic chicane. A great advantage of our method is that in includes no path delay of radiation pulse in the monochromator. Implementation of the proposed technique in baseline undulator will not perturb the baseline mode of operation.
We present an illustration of the scheme for the LCLS-II soft X-ray beam line, although other facilities like FLASH-II, SwissFEL, FERMI and SPARC may also benefit from it.

\section*{Acknowledgements}

We are grateful to Massimo Altarelli, Reinhard Brinkmann, Serguei
Molodtsov and Edgar Weckert for their support and their interest
during the compilation of this work.

\end{document}